\documentclass[letterpaper,12pt]{article}

\usepackage[dvips]{graphicx}  
\usepackage{fullpage}
\usepackage{epsfig}
\usepackage{sectsty}
\newcommand{\ltsimeq}{\raisebox{-0.6ex}{$\,\stackrel
        {\raisebox{-.2ex}{$\textstyle <$}}{\sim}\,$}}

\begin{document}

\sectionfont{\large}
\subsectionfont{\normalsize}

\begin{center}
\Large\bfseries
Black Hole Cross Section at the Large Hadron Collider
\end{center}

\bigskip

\begin{center}
\small
Douglas M. Gingrich \\

\bigskip

\textit{Department of Physics, University of Alberta, Edmonton, AB T6G
2G7 Canada}\\  
\textit{TRIUMF, Vancouver, BC V6T 2A3 Canada}\\ 
{\footnotesize gingrich@ualberta.ca}

\bigskip

\end{center}

\begin{quotation} \noindent
\textbf{Abstract\ } 
Black hole production at the Large Hadron Collider (LHC) was first
discussed in 1999.
Since then, much work has been performed in predicting the black hole
cross section.
In light of the start up of the LHC, it is now timely to review the state
of these calculations. 
We review the uncertainties in estimating the black hole cross section
in higher dimensions. 
One would like to make this estimate as precise as possible since the
predicted values, or lower limits, obtain for the fundamental Planck 
scale and number of extra dimensions from experiments will depend
directly on the accuracy of the cross section.    
Based on the current knowledge of the cross section, we give a range of 
lower limits on the fundamental Planck scale that could be obtained at
LHC energies.
\end{quotation}

\begin{quotation} \noindent
\end{quotation}

\section{Introduction}

Models with large\cite{ADD1,ADD2,ADD3} or warped\cite{RS1,RS2} extra
dimensions allow the fundamental scale of gravity to be as low as the
electroweak scale. 
For energies above the gravity scale, black holes can be produced in
particle collisions.
This opens up the possibility to produce black holes at the Large Hadron
Collider (LHC).
The cross section is approximately given by the horizon area of the
black hole.
Once formed, the black hole can decay by emitting Hawking radiation.
The final fate of the black hole is under much debate: quantum gravity
will be involved and a stable remnant is possible. 
If produced at the LHC, black holes will not only allow us to test
classical gravity and probe extra dimensions, but will also teach us
about quantum gravity.

Over the last seven years there has been much debate over the form of
the black hole production cross section in particle collisions.
Early discussions postulated a $\pi R_\mathrm{S}^2$ form for the cross
section,\cite{Banks,Giddings1,Dimopoulos1} where $R_\mathrm{S}$ is the
Schwarzschild radius of the black hole formed in the hard scattering.  
This naive form has been criticised\cite{Voloshin1,Voloshin2} and
defended\cite{Solodukhin,Jevicki,Rychkov04} over the years. 
It is well known that this simple form does not take into account the
angular momentum of the black hole.
Attempts have been made to account for angular momentum in a heuristic
way by multiplying the simple expression for the cross section by a form 
factor.\cite{Park,Kotwal,Anchordoqui1,Ida} 
Still, recent work most often takes the form factor to be unity. 

Calculations based on classical general relativity have had limited
success in improving the cross section
estimates.\cite{Eardley,Yoshino1}
The effects of mass, spin, charge, and finite size of the incoming
particles are usually neglected. 
The effects of finite size have been discussed,\cite{Kohlprath} and
only recently have angular momentum or charge been
considered.\cite{Yoshino2,Mann}   
Although the calculations are far from complete, they do indicate that
the simple geometric cross section is correct if multiplied by a
formation factor, which has a value at most of about three for seven
extra dimensions. 
These calculations have also been criticised\cite{Rychkov} and defended
over the years.\cite{Giddings2} 

Not all of the available energy in a hard parton collision will be
trapped behind the event horizon of the black hole.
Although it is not known how the energy will be lost, most energy will
probably be lost to gravitational radiation into the bulk during the
collision. 
The effects of trapped energy can become important when calculating the
particle-level cross section since they result in lowering the mass of
the black hole that can be produced for a given hard parton collision.
Although only lower limits on the trapped energy have been
calculated,\cite{Yoshino2} these limits allow us to recalculate the
particle level cross sections with the assumption of trapped energy.
The resulting lower cross sections could modify previous phenomenological
results on black hole production at the LHC.

An outline of this paper is as follows.
In section 2, we clearly state our assumptions and approximations.
This helps define the class of models considered.
A review of the classical parton cross section is given in
section 3.
In section 4, we discuss and write down the particle-level cross section.
A variety of improvements to the classical cross section are discussed
in section 5.
We also summarize cross section results from some other
models. 
In section 6, estimates of the form factor are given.
We take trapped energy into account and give limits on the cross
section in section 7.
And finally in section 8, we derive lower limits on the Planck scale
based on the best estimates of the cross section.
Black hole decays are not discussed in this paper.

\section{Assumptions and Approximations}

Black hole solutions in higher dimensions have a complicated dependence
on both the gravitational field of the brane and the geometry of the
extra dimensions.  
Test particles of a sufficiently high energy to resolve a distance as
small as the Planck length are predicted to gravitationally curve, and
thereby to significantly disturb, the very spacetime structure that they
are meant to probe. 
However, there are two useful approximations that may be used for a wide
class of solutions.\cite{Giddings1}

\begin{enumerate}
\item 
The brane is expected to have a tension given by roughly the Planck or
string scale. 
Moreover, the brane on which the Standard Model particles live, will have
a gravitational field that should be accounted for in solving
Einstein's equations.\cite{Aliev,Rocha}   
For black holes with mass $M$ substantially heavier than the fundamental
Planck scale in higher dimensions $M_D$, $M \gg M_D$, the brane's field
should be negligible and the production process for black holes should
be non-perturbative.    
We will assume that the only effect of the brane field is to bind the
black hole to the brane, and that otherwise the black hole may be
treated as an isolated object in the extra dimensions. 
This is often referred to as the ``probe brane approximation''. 
The effects of quantum gravity will be small under this approximation.

\item 
If the geometrical scales of the extra dimensions $R$ (radii, curvature 
radii, variation scale of the warp factor) are all large compared to
$1/M_D$, then there is a wide regime in which the geometry of the extra
dimensions plays no essential role.   
An alternative view of this condition is to only consider black holes
with horizon radius $r_\mathrm{h}$ much smaller than the size of the
extra dimensions, $1/M_D < r_\mathrm{h} \ll R$, since the Compton
wavelength of a black hole is smaller than its horizon radius for most
of the parameter space. 
For large flat dimensions, the topology of the black hole can be assumed 
to be spherically symmetric in $(n+3)$-spatial dimensions, and the
boundary conditions from the compactification can be neglected.
\end{enumerate}

\noindent
Within this region of applicability, it is often a good approximation to
consider the high-energy collision of the particles and the black hole
formed to be in $(n+4)$-dimensional flat spacetime. 
The above two conditions will be used to define the class of models
under which we will examine the black hole cross section.
In addition, we assume the lifetime of the black hole is long enough so
that it behaves as a well-defined quasi-stable state.
Some alternative forms of the cross section have been derived under
different conditions and will be briefly mentioned. 

\section{Classical Parton Cross Section}

The black hole is defined as any matter or energy trapped behind the
horizon formed by the available mass and energy of the
particle collision.
For a particular amount of available energy, a range of black hole
masses will result depending on the impact parameter of the particle
collision. 
For an impact parameter $b \le r_\mathrm{h}$, the incident relativistic
particles pass within the horizon.
Formation of the horizon should occur before the particles come in
causal contact and would be a classical process.
Once inside the horizon, no matter how strong the subsequent QCD 
effects become, formation of an excited black hole state results. 
The production of a black hole in high energy collisions would be a
totally inelastic process.

In the high-energy limit, in which the classical picture is valid, a
black hole can be formed for any incident center of mass energy.  
The black hole should not be thought of as a single massive degree of
freedom,  but rather as a intermediate ``resonance'' with effectively a
continuum of states representing the large number of black hole masses.

Since the black hole is not an ordinary particle of the Standard Model
and its correct quantum theoretical treatment is unknown, it is treated
as a quasi-stable state, which is produced and decays according to the
semiclassical formalism of black hole physics.
Using the above approximations, it has been argued
that at high energies black hole production has a good classical
description.\cite{Giddings1,Dimopoulos1}  
This leads to the naive estimate that the cross section for black hole
production is approximately given by the classical geometric cross
section

\begin{equation} \label{eq1}
\hat{\sigma} = \pi R_\mathrm{S}^2 ,
\end{equation}

\noindent
where $R_\mathrm{S}$ is the $(n+4)$-dimensional Schwarzschild radius
corresponding to the black hole mass $M$ (see appendix for the explicit
form of the horizon radius).  
It depends on the fundamental Planck scale $M_D$ and the number of extra
dimensions $n$.
In the high-energy limit, the cross section should depend on the impact
parameter $b$, and a range of black hole masses will result for a given
center of mass energy. 
Since the cross section is dominated geometrically by large impact
parameters $b \ltsimeq R_\mathrm{S}$, the average black hole mass should be of
the order of the center of mass energy, $\langle M\rangle \ltsimeq
\sqrt{\hat{s}}$. 
In Eq.~(\ref{eq1}), it is assumed that the black hole mass is given by $M
= \sqrt{\hat{s}}$. 
The expression does not contain any small coupling constants that we
have to compared to perturbative physics processes; the black hole cross
section grows much faster than any known perturbative local physics.   
The expression for the cross section only contains the fundamental
Planck scale as a coupling constant.

In calculating the parton-level cross section and comparing with
previous work, it is important to be aware of the many different
conventions for the Planck scale.
In addition, there are different conventions for the extra-dimensional
Newton constant and the definition of the compactification radius $R$.
Throughout this paper we used the Particle Data Group (PDG)\cite{PDG}
definition of the Planck scale

\begin{equation}
M_D^{n+2} = \frac{1}{8\pi G_\mathrm{N}} \frac{1}{R^n} ,
\end{equation}

\noindent
where $G_\mathrm{N}$ is Newton's constant in four dimensions.
This convention is often referred to as the Giudice, Rattazzi,
Wells\cite{Giudice} or GRW convention.
Another very popular convention that leads to a simple form for the
cross section is given by Dimopoulos and Landsberg\cite{Dimopoulos1}
(DL convention) 

\begin{equation}
M_\mathrm{DL}^{n+2} = \frac{1}{G_\mathrm{N}} \frac{1}{(2\pi R)^n} .
\end{equation}

\begin{figure}[p]
\begin{center}
\includegraphics[width=15cm]{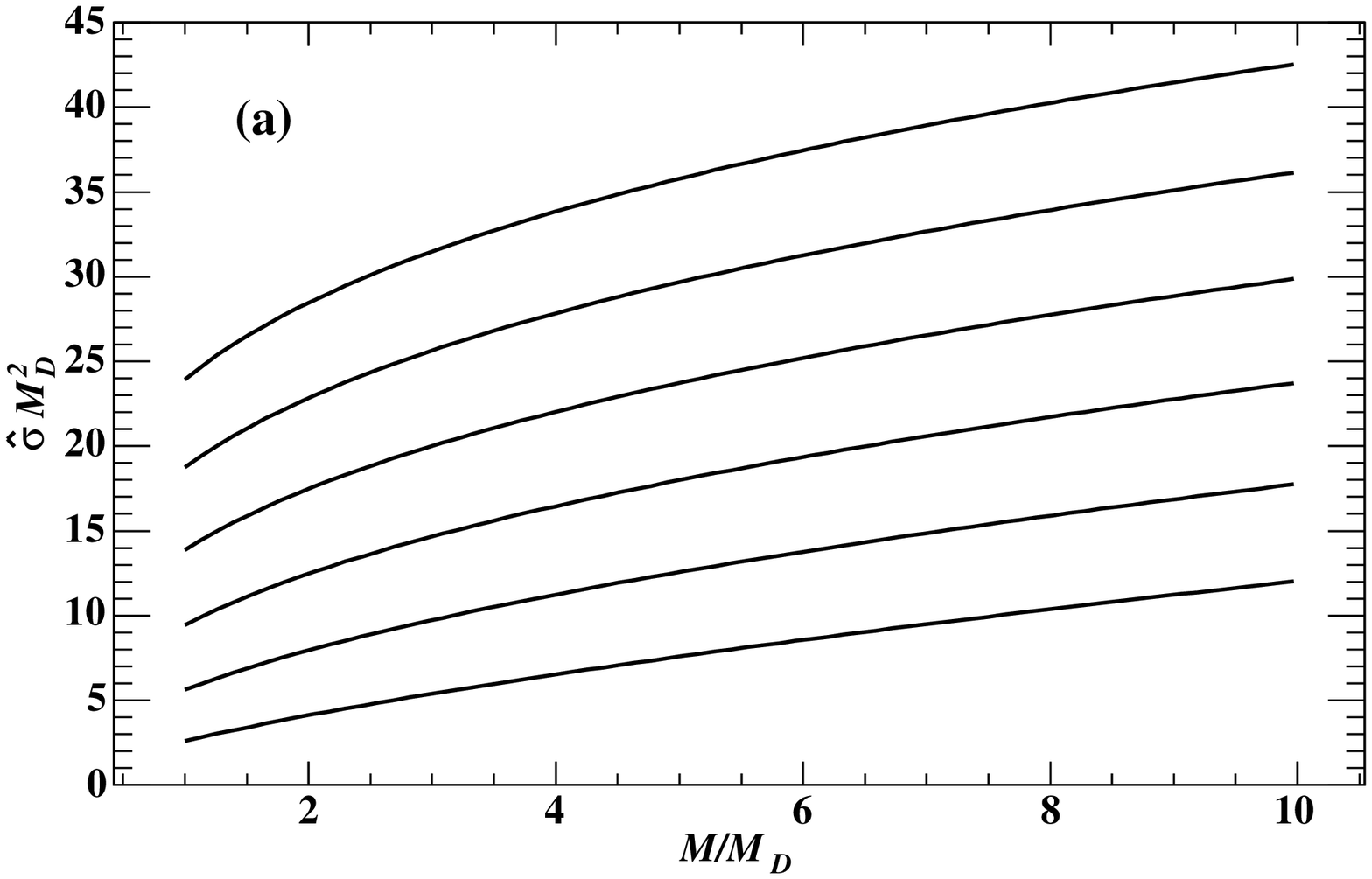}
\hfill
\includegraphics[width=15cm]{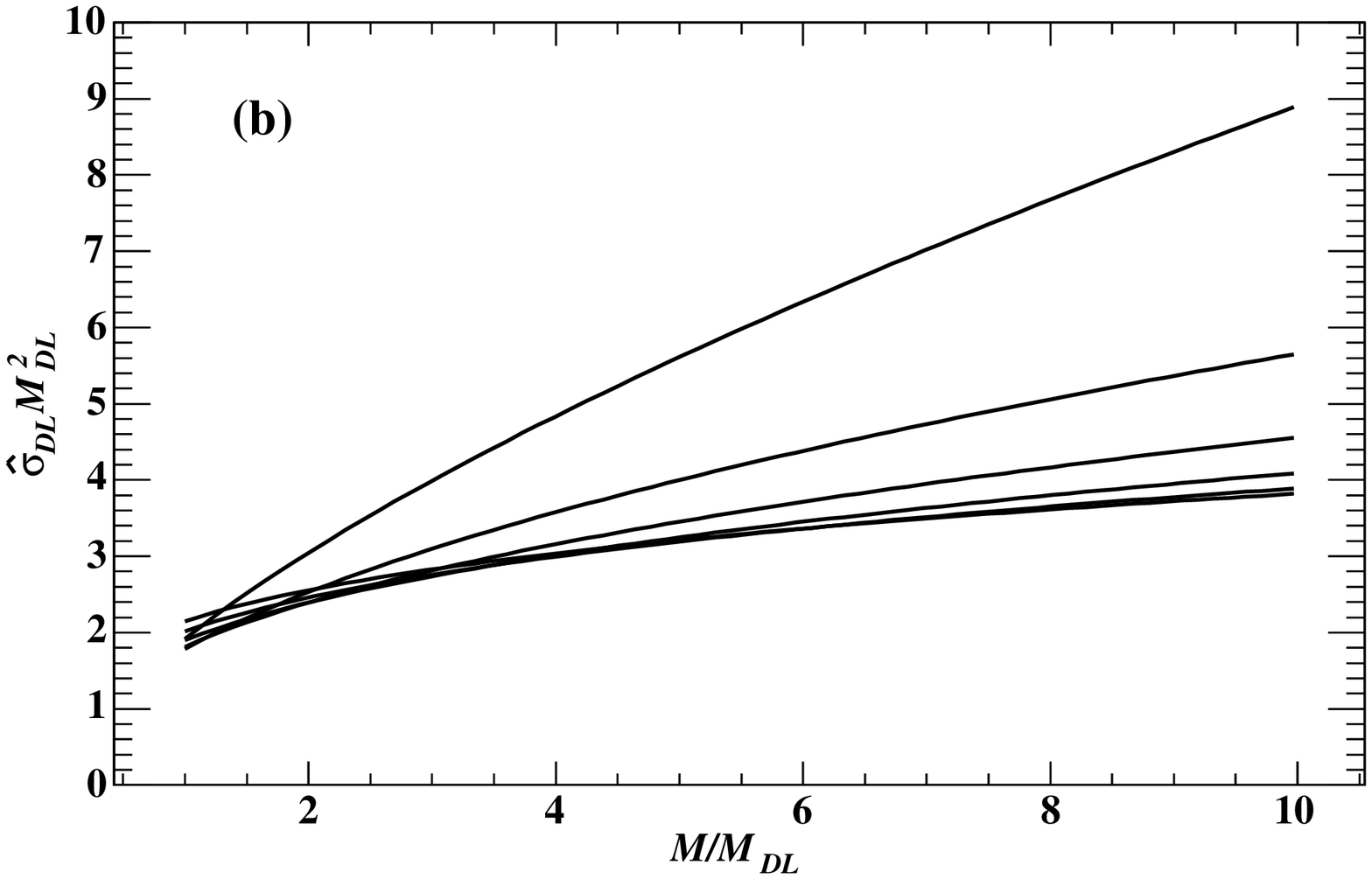}
\caption{\label{part}Parton cross section (units of inverse Planck
scale squared) versus black hole mass (units of Planck scale) for
different $n$. 
(a) PDG definition of Planck scale; $\hat{\sigma}M_D^2$ increases as
$n=2$--7.  
(b) Dimopoulos and Landsberg definition of Planck scale;
$\hat{\sigma}_\mathrm{DL}M_\mathrm{DL}^2$ decreases as $n=2$--7.}
\end{center}
\end{figure}

\noindent
Although differences in the conventions can be ignored for
astrophysical or cosmological work, they are important and must be taken
into account at LHC energies. 
The DL convention has a weaker dependence on $n$ than the PDG convention. 
For $M_D = M_\mathrm{DL}$, the cross section given by the PDG convention
is greater than that of the DL convention for $n \ge 2$ (see
appendix)\cite{Rizzo1}.  
To illustrate the effect the definition of the Planck scale can
have, we plot the parton cross section versus black hole mass in
Fig.~\ref{part}(a) and \ref{part}(b) using the PDG and DL conventions
for the Planck scale, respectively. 
We plot dimensionless quantities for the cross section and mass to
emphasis the differences due to the Planck scale conventions. 
The effect of the Planck scale definition on the dimensionless cross
section is significant for most black hole masses and number of extra
dimensions. 

\section{Particle Cross Section}

Only a fraction of the total center of mass energy $\sqrt{s}$ in a
proton-proton collision is available in the hard scattering process.
We define

\begin{equation}
s x_\mathrm{a} x_\mathrm{b} \equiv s \tau \equiv \hat{s} ,
\end{equation}

\noindent
where $x_\mathrm{a}$ and $x_\mathrm{b}$ are the fractional energies of
the two partons relative to the proton energies.
The full particle-level cross section $\sigma$ is obtained from the
parton-level cross section $\hat{\sigma}$ by using

\begin{equation} \label{eq3}
\sigma_{pp\to \mathrm{BH+X}}(s) = \sum_\mathrm{a,b} \int^1_\frac{M^2}{s}
dx_\mathrm{a}
\int^1_\frac{M^2}{x_\mathrm{a}s} dx_\mathrm{b}
f_\mathrm{a}(x_\mathrm{a})
f_\mathrm{b}(x_\mathrm{b})
\; \hat{\sigma}_\mathrm{ab\to BH}(\hat{s}=M^2) ,
\end{equation}

\noindent
where a and b are the parton types in the two protons, and
$f_\mathrm{a}$ and $f_\mathrm{b}$ are parton distribution functions
(PDFs) for the proton. 
The sum is over all possible quark and gluon pairings.
The parton distributions fall rapidly at high relative energies, and so
the particle-level cross section also falls at high energies.

The momentum scale $Q$ at which the parton distribution functions are
evaluated is determined by the inverse length scale associated with the
scattering process. 
For perturbative hard scattering in a local field theory this momentum
scale is given by the momentum transfer, which in the $s$-channel is the
parton-parton center of mass energy: $Q \sim \sqrt{\hat{s}}$.
For the non-perturbative process of $s$-channel black hole formation in
a theory of classical gravity, the relevant length scale is the horizon, 
rather than the black hole mass: $Q \sim
R_\mathrm{S}^{-1}$~\cite{Emparan}.  
This makes sense since the size of the black hole horizon is bigger than
its Compton wavelength.  

Throughout this paper we use the CTEQ6L1 (leading order with leading
order $\alpha_s$) parton distributions functions\cite{Pumplin} within
the Les Houches Accord PDF framework\cite{LHAPDF}.
We have taken $Q = R_\mathrm{S}^{-1}$ for the scale.

Making a change of variables from $(x_\mathrm{a},x_\mathrm{b})$ to
$(\tau,x)$ in Eq.~(\ref{eq3}) gives 

\begin{equation}
\sigma_{pp\to \mathrm{BH+X}}(s) = \sum_\mathrm{a,b} \int^1_\frac{M^2}{s}
d\tau \int^1_\tau \frac{dx}{x} f_\mathrm{a}\left(\frac{\tau}{x}\right)
f_\mathrm{b}(x) 
\; \hat{\sigma}_\mathrm{ab\to BH}(\hat{s}=M^2) .
\end{equation}

\noindent
The differential cross section can be written as

\begin{equation}
\frac{d\sigma_{pp\to \mathrm{BH+X}}}{d\tau}(s) = \sum_\mathrm{a,b}
\int^1_\frac{M^2}{s} \frac{dx}{x} f_\mathrm{a}\left(\frac{\tau}{x}\right)
f_\mathrm{b}(x) \nonumber \\
\; \hat{\sigma}_\mathrm{ab\to BH}(\hat{s}=M^2) .  
\end{equation}

\noindent
Since $\hat{s} = M^2$, we can make a changing of variable from $\tau$ to
$M$, $dM/d\tau = s/(2M)$, to obtain

\begin{equation}
\frac{d\sigma_{pp\to \mathrm{BH+X}}}{dM}(s) = \frac{2M}{s}
\sum_\mathrm{a,b} \int^1_{M^2/s} \frac{dx}{x}
f_\mathrm{a}\left(\frac{\tau}{x}\right) f_\mathrm{b}(x) \nonumber \\
\; \hat{\sigma}_\mathrm{ab\to BH}(\hat{s}=M^2) .  
\end{equation}

\noindent
In terms of parton luminosity (or parton flux), we write

\begin{equation} \label{eq7}
\frac{d\sigma_{pp\to \mathrm{BH+X}}}{dM} = \frac{dL}{dM}\;
\hat{\sigma}_\mathrm{ab\to BH} ,
\end{equation}

\noindent
where

\begin{equation} \label{eq8}
\frac{dL}{dM} = \frac{2M}{s} \sum_\mathrm{a,b} \int^1_{M^2/s}
\frac{dx}{x} f_\mathrm{a}\left( \frac{\tau}{x} \right) f_\mathrm{b}(x) .
\end{equation}

The differential cross section thus factorizes for the case of $\hat{s}
= M^2$.
It can be written as the product of the parton cross section time a
luminosity function. 
The parton cross section is independent of the parton types and depends
only on the black hole mass, Planck scale, and number of extra
dimensions.
The parton luminosity function contains all the information about the
partons. 
Beside a dependence on black hole mass, it is independent of the
characteristics of the higher-dimensional space, i.e.\ the Planck scale
and number of extra dimensions.  
The dependence on the black hole mass occurs in the proportionality, the
limit of integration, and in the scale of the parton density functions.
If the horizon is used as the distance scale in the parton density
functions, the luminosity function will depend indirectly on the Planck
scale and number of extra dimensions.
 
\begin{figure}[tb]
\begin{center}
\includegraphics[width=15cm]{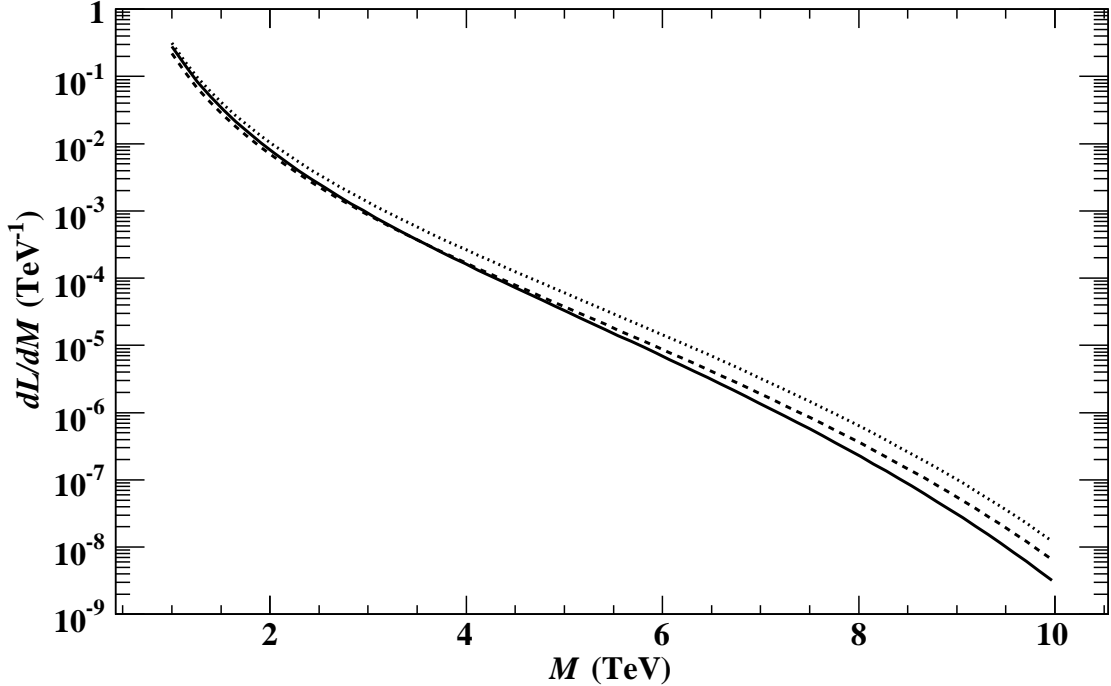}
\caption{\label{lumi}Parton luminosity versus black hole mass. 
Solid curve $Q = M$, dashed curve $Q = R_\mathrm{S}^{-1}$ with $M_D =
1$~TeV and $n=0$, and dotted curve $Q = R_\mathrm{S}^{-1}$ with $M_D =
1$~TeV and $n=7$.}   
\end{center}
\end{figure}

For a fixed proton-proton center of mass energy, the parton luminosity
function can be pre-calculated to obtain a function depending only on
the single mass parameter.  
Figure~\ref{lumi} shows the parton luminosity function versus black hole 
mass for $\sqrt{s} = 14$~TeV for different choices of the QCD scale in
the parton density functions for the proton. 
The choice of scale is clearly significant at high black hole masses.
Thus the particle-level cross section does not truly factorize if the
horizon radius is used as the QCD scale in the parton density
functions for the proton.
The steep decrease in Fig.~\ref{lumi} with black hole mass kills any rise
in the parton cross section with black hole mass.
Nevertheless, if the Planck scale is at the TeV level, particle
scattering in the $s$-channel will be dominated by black hole production.

The transition from the parton-level to the hadron-level cross section
is based on a factorization formula.
The validity of this formula for the energy region above the Planck
scale is unclear.
Even if factorization is valid, the extrapolation of the parton
distribution functions into this transplanckian region based on
Standard Model evolution from present energies is questionable, since
the evolution equations neglect gravity. 
To proceed, we ignore these difficulties.

\section{Improvements to the Cross Section}

In studying the uncertainties in Eq.~(\ref{eq1}), it is useful to examine
a more general form of the cross section

\begin{equation} \label{eq2}
\sigma = F \pi r_\mathrm{h}^{2+k}\Theta(M - M_\mathrm{min}) ,
\end{equation}

\noindent
where $F$ is a form factor (usually approximated as unity),
$r_\mathrm{h}$ is a more general horizon that may depend on the angular
momentum and charges of the black hole (usually taken to be the
non-spinning non-charged Schwarzschild radius in $(n+4)$-dimensions),
$k$ is the number of extra dimensions in which the Standard Model
particles can propagate (usually  taken as $k=0$), and $\Theta$ is a
Heaviside step function that allows black hole production only above
some threshold mass $M_\mathrm{min}$ (often implicitly assumed).  
Implicit in Eq.~(\ref{eq2}) is the possibility that each of the factors 
may depend on the black hole mass, angular momentum, and charges, as well
as the fundamental Planck scale and number of extra dimensions.
In addition, $M < \sqrt{\hat{s}}$ needs to be considered to allow for
the possibility of not all the available energy being trapped behind the
horizon.  
We now comment on each of the possible modifications to Eq.~(\ref{eq1})
in Eq.~(\ref{eq2}).

\subsection{Form Factor}

Semiclassical considerations give rise to a form factor that reflects
the theoretical uncertainties in the dynamics of the process. 
They take into account that not all of the initial center of mass energy
is captured behind the horizon, or account for the distribution of black
hole mass as function of energy and angular momentum.
In general, the form factor will depend on the parton center of mass
energy, the resulting black hole mass, and angular momentum, as well as
the fundamental Planck scale and number of extra dimensions.
Most studies take the numerical value of the form factor to be unity but
we suggest it be accounted for by using the results given below.

\subsection{Horizon Radius}

An intermediate resonance produced in a parton-parton collision must
carry the gauge and angular momentum quantum numbers of the initial
parton pair. 
In the high-energy limit, black hole states exist with gauge and angular
momentum quantum numbers corresponding to any possible combination of
partons within the proton.
Thus spinning black holes are expected to be produced at colliders.
Although qualitatively equivalent to non-spinning black holes, the
results for spinning black holes are expected to be quantitatively quite
different and probably more realistic. 
The Kerr solutions for a spinning black hole in $(n+4)$-dimensional flat 
spacetime have been calculated\cite{Myers} (see appendix).
The horizon thus depends on the black hole mass and angular momentum, as
well as the fundamental Planck scale and number of extra dimensions.

In general, the Kerr solutions should be used for $r_\mathrm{h}$ but a
single functional form for all dimensions does not exist.
It is usual to make a heuristic argument about the relationship between
the angular momentum and horizon radius or maximum impact parameter, and
then to obtain an approximate expression for the horizon radius in all
dimensions. 
It is convenient to write this expression as the Schwarzschild radius
times an angular momentum-dependent form factor (see appendix).

In the following, we will ignore that the black hole could have charges
depending on the partons involved in the hard collision.
Such considerations could give rise to the exciting possibility of a
naked singularity or a black hole remnant.  
Much attention has been given to the decay process but very little to
the production process of a charged black hole\cite{Koch}.
However, Yoshino and Mann\cite{Mann} have recently examined head-on
collisions of ultra-relativistic charges.
They boosted the Reissner-Nordstr{\"o}m spacetime to the speed of
light. 
Using the slice at the instant of collision, they studied formation of 
the apparent horizon and derived a condition indicating that a critical
value of the electric charge is necessary for formation to take place.
They showed that the presence of charge could decreases the black hole
production rate at the LHC.
Since this work is very preliminary, we consider the topic of charge to
be outside the class of models that we are considering. 

Other interesting models, based on assumptions outside of those
considered here, have been put forward to obtain distinct forms of the
horizon. 
Rocha and Coimbra-Ara{\'u}jo\cite{Rocha} have recently solved
Einstein's equations on the brane to derive the exact form of the
brane-world-corrected perturbations in Kerr-Newman metric
singularities. 
Rizzo\cite{Rizzo2} has performed an analysis in the Randall-Sundrum
model with an extended action containing Gauss-Bonnet terms.
He obtained expressions for the black hole production cross section in
this model.
As in the flat scenario, the cross sections are large.
A mass threshold below which the black hole will not be produced comes
about from a restriction on the Hawking temperature being positive.
This gives the cross section a steplike behaviour near $M_D$. 
Rizzo\cite{Rizzo3} has also examined the modifications to black hole
properties due to the existence of spacetime non-commutativity in string
theory. 
In some cases, these models can give significant modifications to the
cross section.

\subsection{Transmission in the Bulk}

There are scenarios, such as fat branes or universal extra
dimensions\cite{Dimopoulos2,Casadio}, in which the Standard Model
particles are allowed, to some extent, to propagate in the space
dimensions other than the normal $(3+1)$-dimensions.  
If gauge bosons are allowed to propagate in the extra dimensional
spacetime, their interactions would be modified beyond any acceptable
phenomenological limits unless the size of the extra dimensions are
very small. 
Such models can give rise to string balls and $p$-branes, as well as black
holes.
The cross section for producing black holes will be modified and can be
substantial larger than that given in
Eq.~(\ref{eq1}).\cite{Ahn1,Cheung1,Cheung2}  
In scenarios in which the Standard Model particles also propagate in the 
extra dimensions, the incoming partons must collide with an impact 
parameter in the extra dimensions, which is less than $r_\mathrm{h}$ in
all dimensions in order to produce a black hole.  
Therefore, the cross section for producing a black hole must scale as
$r_\mathrm{h}^{2+k}$, where $k$ is the number of extra dimensions in
which the Standard Model particles propagate, not necessarily equal
to $n$ the total number of extra dimensions.

The power of 2 in Eq.~(\ref{eq1}) is the result of considering the
Standard Model particles to propagate on a 3-brane.
In almost all studies of black hole production at colliders this
assumption is made.
We will thus consider this assumption to define our model and will put
$k=0$ in Eq.~(\ref{eq2}).
However, we point out that models allowing for non-vanishing $k$ could
give similar signatures to black hole production and thus result in
the nonsensical experimental determination of the fundamental Planck
scale and the number of extra dimensions if ignored.

Another interesting result arises from split fermions models.
In split fermion models, most problems with TeV-scale gravity can be
solved if different fermions are localized at different points in the
extra dimensions.\cite{Arkani1,Arkani2} 
Dai, Starkman, and Stojkovic\cite{Dai} have examined black holes and
their angular momentum distribution in models with split fermions.
They find that the total production cross section is reduced compared
with models where all the fermions are localized at the same point in
the extra dimensions.

\subsection{Mass Threshold}

There exists a threshold for black hole production.
In classical general relativity, two point-like particles in a head-on
collision with zero impact parameter will always form a black hole, no
matter how large or small their energy.
At small energies, we expect this to be impossible due to the smearing
of the wave function by the uncertainty relation.
This then results in a necessary minimal energy to allow the required
close approach.
The threshold is of order $M_D$, though the exact value is unknown since
quantum gravity effects should play an important role for the wave
function of the colliding particles.
Such a threshold arises naturally in certain types of higher order
curvature gravity.\cite{Hossenfelder}
For simplicity, it is usual to set this threshold equal to $M_D$.

In the high-energy limit, if the impact parameter is less than
$r_\mathrm{h}$, a black hole with mass $M \sim \sqrt{\hat{s}}$ can be
produced.  
To avoid quantum gravity effects and stay in the classical regime, we
require $M \ge M_\mathrm{min}$, where $M_\mathrm{min}$ should be a few
times larger than $M_D$, although it is often taken as $M_D$. 
A reasonable criterion for $M_\mathrm{min}$ is given by the requirement
of large entropy.\cite{Anchordoqui2}
In the following, we will find it useful to define the dimensionless
parameter

\begin{equation}
x_\mathrm{min} = \frac{M_\mathrm{min}}{M_D} ,
\end{equation}

\noindent
and require $x_\mathrm{min} \gg 1$.
Unfortunately, all the results will now depend on the subjective choice of
the $x_\mathrm{min}$ cutoff.

\subsection{Trapped Energy}

Classical general relativistic calculations indicate that the mass of a
black hole formed in a head-on collision is somewhat less than the total
center of mass energy; the scattering is not completely inelastic.
This is because gravitational radiation is expected to dominate, and
because the energy-momentum multipole moments generated during the
process of formation have values within the Standard Model
brane.\cite{Giddings1}  
Thus Eq.~(\ref{eq1}) should be modified by replacing the black hole mass
by a fraction of the available center of mass energy, leading to a
reduction in the cross section.   
Suggestions for how to treat trapped energy are given below.

\subsection{Exponential Suppression and Quantum Gravity}

Models introducing an exponential suppression of the classical cross
section have been formulated.
Voloshin\cite{Voloshin1,Voloshin2} suggested an exponential suppression
of the geometric cross section based on a
Gibbons-Hawking\cite{Gibbons} action argument.  
Detailed subsequent studies performed in simple string theory models,
using full general relativistic calculations or a path integral approach, 
do not confirm this finding, and prove that the geometric cross
section is modified only by a numerical factor of order one, at least up
to energies of about $10 M_D$. 
Solodukhin\cite{Solodukhin} applied a consistent treatment of both
the path integral and statistical approaches suggested in
Ref.~\cite{Voloshin1}, and found no such exponential suppression. 
A flaw in the Gibbons-Hawking action argument was further found.
The use of this action implies that the black hole has already formed,
so describing the evolution of the two colliding particles before they 
cross the horizon and form the black hole.\cite{Jevicki}
The most direct undoing of Voloshin's argument has been made by
Rychkov\cite{Rychkov04}, which resulting in an erratum by
Voloshin\cite{Voloshin3}. 
It has further been shown by Rizzo\cite{Rizzo1} that the black hole
cross section, even with the Voloshin suppression factor, can be large.
We consider the debate to be resolved, and will not consider a Voloshin
exponential suppression factor. 

Even if a full description of quantum gravity is not yet available,
there are some general features reappearing in most candidates for such
a theory: the need for higher dimensions and the existence of a minimal
length scale.
String theory, as well as non-commutative quantum mechanics, suggest that
the Planck length acts as a minimal length in nature, providing a
natural ultraviolet cutoff and a limit to the possible resolution of
spacetime.
The minimal-length effects thus become important in the same energy
range in which the black holes are expected to form.

The influence of minimal-length scale on the production of black holes
in a model with large extra dimensions was examined by
Hossenfelder.\cite{Hossenfelder} 
The finite resolution of spacetime, which is caused by the minimal
length, results in an exponential suppression of black hole production. 
At LHC energies, the total cross section is about a factor of five
smaller under this scenario.
While this is an interest scenario, we will not address aspects of
quantum gravity in this paper.

\section{Estimates of the Form Factor}

Based on the above comments, we now take the black hole cross section to
be given by 

\begin{equation} \label{eq13}
\hat{\sigma} = F \pi R_\mathrm{S}^2\Theta(x_\mathrm{min} -1) .
\end{equation}

\noindent
In the following, we show that the Kerr solution along with heuristic
angular momentum arguments lead to an $n$ dependent form factor times
the Schwarzschild radius, and that general relativistic calculations
likewise lead to similar results.

\subsection{Heuristic Angular Momentum Arguments}

Due to the conservation of angular momentum, the angular momentum $J$ of
the formed black hole only vanishes completely for central collisions
with zero impact parameter.
In the general case for impact parameter $b$, there will be an angular
momentum $J = 2(b/2)(\sqrt{\hat{s}}/2) = b\sqrt{\hat{s}}/2$.  
The black hole will typically be formed with large angular momentum
components.
Since the impact parameter is only non-vanishing in directions along the
brane, the angular momentum lies within the brane directions.
The direction of the angular momentum axis within the Standard Model
brane is perpendicular to the collision axis in the high-energy limit. 

One may improve estimates of the cross section by taking into account
the angular momentum dependence of the horizon radius.
Park and Song\cite{Park} made an early attempt to incorporate angular
momentum into the cross section.
They assumed that the semiclassical reasoning for the non-rotating black
hole was still valid for rotating black holes.
They replaced the Schwarzschild radius by the Kerr solution, and said the
total cross section is the sum of the individual spin cross sections up
to the maximum possible spin $M R_\mathrm{S}$:  

\begin{equation}
\hat{\sigma} = \sum_{J=0}^{MR_\mathrm{S}} \hat{\sigma}(J) =
\sum_{a=0}^\frac{1+n}{2} \left( \frac{1}{1+a^2} \right)^\frac{2}{1+n}
\hat{\sigma}(J=0) , 
\end{equation}

\noindent
where $a = (2+n)J/(2MR_\mathrm{S})$.
The individual spin cross sections $\hat{\sigma}(J>0)$ are smaller than
the non-spin cross section, and they decrease with increasing spin; this
behaviour is counter intuitive.
However, the total cross section obtained by summing the rotating and
non-rotating cross sections is about 2 to 3 times higher than the
non-rotating case.

Kotwal and Hays\cite{Kotwal} have analyzed the angular momentum
distribution of black holes by computing the production probability
using the partial wave expansion of the initial state.
They assumed a step-function interaction Hamiltonian with an arbitrary 
normalization. 
Two choices for the phase space were examined: purely geometric (no
phase space) and number of available states given by the entropy.
Fairly different results were obtained depending on the phase space
used.
No overall normalization was provided and nor was a closed-form
expression obtained for the cross section.
We find this model of little practical use for making predictions at the
LHC.

One expects the maximum impact parameter will occur near a value of $b$
that equals the corresponding angular momentum dependent radius
$r_\mathrm{h}$. 
Using the radius of a Kerr black hole and substituting $J =
r_\mathrm{h}M/2$, Anchordoqui \textit{et al}.\cite{Anchordoqui1} obtained  

\begin{equation}
F(n) = \left[ 1 + \frac{(n+2)^2}{16} \right]^\frac{-2}{n+1} .
\end{equation}

\noindent
This result give $F = 0.63$ to 0.64 for $n = 0$ to 7.
This correction is approximately a constant of order unity.

An improved heuristic argument has been given by Ida, Oda, and
Park.\cite{Ida} 
A black hole is formed when

\begin{equation}
b \le 2r_\mathrm{h}(\sqrt{\hat{s}},J) = 2r_\mathrm{h}(M, bM/2) .
\end{equation}

\noindent
Since the right hand side is a monotonically decreasing function of $b$,
there is a maximum value $b_\mathrm{max}$ that saturates the inequality.
When $b=b_\mathrm{max}$, the rotation parameter takes the maximal value,
and one obtains

\begin{equation} \label{eq17}
F(n) = 4 \left[ 1 + \left( \frac{n+2}{2} \right)^2
  \right]^\frac{-2}{n+1} .
\end{equation}

\noindent
The results using this form factor are given in the row labeled
$F_\mathrm{IOR}$ in Table~\ref{t1}.
This correction increases the cross section at most by 1.9 for $n=7$.

Since the element of impact parameter $[b,b+db]$ contributes to the
cross section an amount $2\pi bdb$, the relationship between $b$ and $J$
gives us the differential cross section of a black hole with element of
angular momentum $[J,J+dJ]$:  

\begin{equation} \label{eq18}
\frac{d\hat{\sigma}}{dJ} = \frac{8\pi J}{\hat{s}}
\Theta(J_\mathrm{max}-J) ,
\end{equation}

\noindent
where $J_\mathrm{max} = M R_\mathrm{S}$.
The differential cross section linearly increases with angular momentum.
We expect that this behaviour is correct as a first approximation, so
that black holes tend to be produced with large angular momentum.
Integrating Eq.~(\ref{eq18}) gives Eq.~(\ref{eq17}) times the classical
cross section. 

We point out that these heuristic angular-momentum arguments only
consider the angular momentum of the initial partons; they neglect the
spin of the partons.  

\subsection{General Relativistic Analytical Solutions}

To improve the naive picture of colliding point particles, we need to
consider the grazing collision of particles in $(n+4)$-dimensional Einstein
gravity and investigate the formation of apparent horizons.   
A common approach is to treat the creation of the horizon as a collision
of two shock fronts in Aichelburg-Sexl geometry.\cite{Aichelburg}
The Aichelburg-Sexl metric is obtained by boosting the Schwarzschild
metric to form two colliding shock fronts.
It is assumed that the shock waves can be boosted to thin fronts, thus
neglecting the uncertainty of the quantum particles. 
This treatment is justified as the particles with energy $\sqrt{\hat{s}}
> M_D$ have a position uncertainty smaller than their horizon.
Due to the high velocity of the moving particles, space time before and
after the shocks is almost flat and the geometry can be examined for the
occurrence of trapped surfaces,\cite{Penrose,Eath1,Eath2,Eath3} which
depend on the impact parameter.

Eardley and Giddings\cite{Eardley} developed a method for finding the
apparent horizons for this system.
For a nonzero impact parameter, they were only able to solve the problem
analytically for the $n=0$ case.
They obtained a lower limit of $F(n=0) > 0.65$. 
This result agrees well with the results obtained by Anchordoqui
\textit{et al}.\cite{Anchordoqui1} based on the heuristic spin
arguments above.  

Eardley and Giddings also obtain limits on the final mass of the black
hole formed in $n=0$. 
They found a range from $M > 0.71\sqrt{\hat{s}}$ for $b=0$ to $M >
0.45\sqrt{\hat{s}}$ for $b=b_\mathrm{max}$. 
This can be compared with a perturbative analysis that gave $M
\approx 0.8\sqrt{\hat{s}}$.\cite{Eath1,Eath2,Eath3}
For higher dimensions, they only solved the $b=0$ case to obtained lower 
bounds on the final black hole mass of $M > 0.71\sqrt{\hat{s}}$ to
$0.589\sqrt{\hat{s}}$ for $n=0$ to 7.  

Unfortunately the Eardley and Giddings results are not general enough to
be useful for nonzero impact parameters and higher dimensions.
The results do suggest the form factors are of order unity and the
semiclassical cross section is valid.
They also indicate that a significant amount of the initial energy may
not be trapped behind the horizon.
For more general results, we now turn to the numerical solutions.

\subsection{General Relativistic Numerical Solutions}

Understanding the case of a nonzero impact parameter in higher
dimensions is crucial to improving the cross section estimates.
The analytic techniques used to study head-on collisions in general
relativity are not applicable to collisions at nonzero impact
parameter. 
Thus the claim that a black hole will be produced when $b <
r_\mathrm{h}$ can only be expected to be true up to a
numerical factor. 

Yoshino and Nambu\cite{Yoshino1} solved this problem numerically for
$n>0$ and obtained the maximal impact parameter $b_\mathrm{max}$.
They found that the formation of an apparent horizon occurs when the
distance between the colliding particles is less than 1.5 times the
effective gravitational radius of each particle.
Form factors were obtained and are shown as the row labeled
$F_\mathrm{YN}$ in Table~\ref{t1}.   
The estimated numerical errors in $F_\mathrm{YN}$ are less than 0.4\%
for all values of $n$.
The analytical results of Ida, Oda, and Park ($F_\mathrm{IOP}$) match
the numerical results by Yoshino and Nambu ($F_\mathrm{YN}$), within an 
accuracy of less than 1.5\% for $n\ge 2$ and 6.5\% for $n=1$. 

In their analysis, the apparent horizon was constructed on the union of
the two incoming shocks: ``old slice''. 
However, this slice is not at all optimal in the sense that there exists
other slices located in the future of the old slice. 
Yoshino and Rychkov\cite{Yoshino2} improved the analysis by using the
optimal slice. 
They have updated the calculation and find a 40\% to 70\% increases in
the higher-dimensional cross sections. 
They have also presented rigorous lower bounds on the final irreducible
mass of the black hole, and contours of black hole angular momentum
versus mass. 
The ultimate goal would be to derive a differential cross section
depending on mass and angular momentum of the black hole produced for a
given $\sqrt{\hat{s}}$.  
The resulting form factors are shown as the row labeled $F_\mathrm{YR}$
in Table~\ref{t1}.   

The validity of the setup of the high-energy two-particle system
described above has been questioned because of large-curvature effects 
in the collision of shock waves.\cite{Rychkov,Cardoso1}
However, Giddings and Rychkov\cite{Giddings2} have shown that the
objections were an artifact of the unphysical classical point-particle
limit, and that for a particle described by a small quantum wave packet,
large curvatures do not arise.

\begin{table}[htb] 
\begin{center}
\begin{tabular}{|c|cccccccc|} \hline
$n$ & 0 & 1 & 2 & 3 & 4 & 5 & 6 & 7 \\ \hline
$F_\mathrm{IOP}$ & 1.000 & 1.231 & 1.368 & 1.486 & 1.592 & 1.690 & 1.780
& 1.863 \\ 
$F_\mathrm{YN}$ & 0.647 & 1.084 & 1.341 & 1.515 & 1.642 & 1.741 & 1.819
& 1.883 \\ 
$F_\mathrm{YR}$ & 0.71 & 1.54 & 2.15 & 2.56 & 2.77 & 2.95 & 3.09 & 3.20 \\
\hline 
\end{tabular}
\caption{Form factor $F$ versus the numbers of extra dimensions $n$
for different calculations.
$F_\mathrm{IOP}$ is from reference~\cite{Ida}, $F_\mathrm{YN}$ from
reference~\cite{Yoshino1} and $F_\mathrm{YR}$ from reference~\cite{Yoshino2}.}
\label{t1}
\end{center}
\end{table}

\section{Cross Section at the LHC}

From the above discussion, we recommend three corrections to the
classical cross section: 

\begin{enumerate}
\item using form factors,
\item allowing for non-trapped energy, and
\item applying a minimum black hole mass cutoff. 
\end{enumerate}

\noindent
For the form factors, we recommend using Eq.~(\ref{eq13}) with the values
of $F_\mathrm{YR}$ given in Table~\ref{t1}.
Although these form factors disagree with the other results in
Table~\ref{t1}, they are based on the most detailed calculation to
date. 
It would appear that form factors increase the cross section by a
factor of a few for large $n$, but are irrelevant for order-of-magnitude 
estimates.
The following describes one approach to taking estimates of the
non-trapped energy into account, and applying a minimum black hole mass
cutoff to final results.

\subsection{Trapped Energy Estimates}

\begin{figure}[tb]
\begin{center}
\includegraphics[width=15cm]{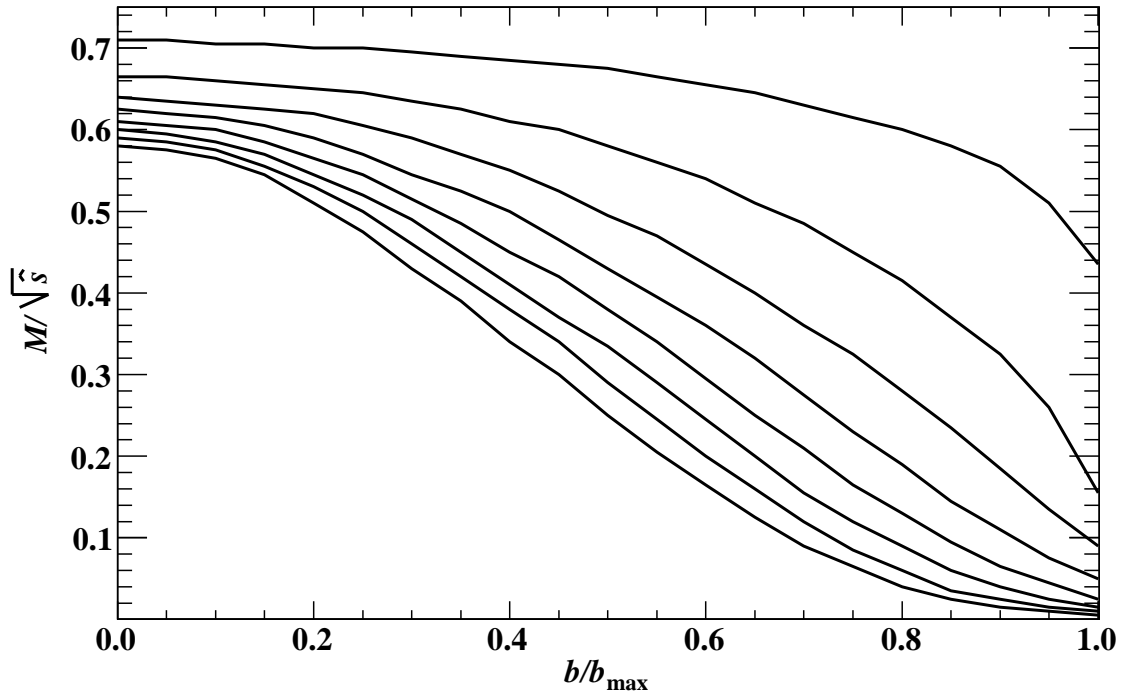}
\caption{\label{trapped}Fractional trapped energy versus scaled impact
parameter for different $n$.    
Top curve $n=0$ and bottom curve $n=7$.  
Data taken from Ref.~\protect\cite{Yoshino2}.}  
\end{center}
\end{figure}

Yoshino and Nambu\cite{Yoshino1} (updated by Yoshino and
Rychkov\cite{Yoshino2}) have provided rigorous lower bounds on the
amount of available energy trapped behind the horizon.
We use their data to obtain lower bounds on the black hole cross
section. 
Figure~\ref{trapped} shows the minimum fraction of energy going into the
production of the black hole versus scaled impact parameter for
different numbers of extra dimensions.
These are lower limits and do not depend on the fundamental Planck
scale. 

We see that the lower bound on the black hole mass formed is never more
than 71\% of the available energy.
The fraction of energy available decrease with impact parameter and the
number of extra dimensions, from 0.71 to 0.46 for $n=0$ to 0.59 to 0 for
$n=7$. 
The mean lower bound on the trapped energies are about 0.6 and 0.27 for
$n=0$ and 7 respectively. 
In higher-dimensional spacetime, the amount of ``junk'' energy
increases because the gravitational field distributes in the space of
the extra dimensions and only a small portion of the total energy of the
system can contribute to the horizon formation.
This junk energy will be radiated away rapidly, probably by
gravitational radiation in the bulk.

\subsection{Particle Cross Section}

Previous calculations of the cross section for producing a black hole
have neglected energy loss in the creation of a black hole, and assumed
that the mass of the created black hole was identical to the incoming
parton center of mass energy.
However, recent work\cite{Yoshino1,Yoshino2} shows the energy loss
to gravitational radiation is not negligible, and in fact is large for 
large number of extra dimensions and for large impact parameters (see
Fig.~\ref{trapped}).  

The trapped mass $M$ is given by (using the notation of Anchordoqui
\textit{et al}.\cite{Anchordoqui3}) 

\begin{equation}
M(z) = y(z) \sqrt{\hat{s}} ,
\end{equation}

\noindent
where the inelasticity $y$ is a function of $z \equiv b/b_\mathrm{max}$.
This complicates the parton model calculations, since the production of
a black hole of mass $M$ is lower than $\sqrt{\hat{s}}$ by $M/y(z)$, thus
requiring the lower cutoff on the parton momentum fraction to be a
function of the impact parameter.
We can no longer use the factorized version of the particle-level cross
section given by Eqs.~(\ref{eq7}) and (\ref{eq8}).

Following Anchordoqui \textit{et al}.\cite{Anchordoqui2,Anchordoqui3},
we take the proton-proton cross section as the impact parameter-weighted
average over parton cross sections, with the lower parton fractional
momentum cutoff determined by the requirement $M_\mathrm{min} =
x_\mathrm{min} M_D$. 
This gives a lower bound $(x_\mathrm{min}M_D)^2 / (y^2s)$ on the parton
momentum fraction $x$. 
With this in mind, the $pp\to \mathrm{BH+X}$ cross section becomes 

\begin{equation} \label{eq19}
\sigma_{pp\to \mathrm{BH+X}}(s,x_\mathrm{min}) \ge \int^1_0 2zdz
\sum_\mathrm{a,b} \int^1_{\frac{(x_\mathrm{min}M_D)^2}{y^2s}}
d\tau
\; \int^1_\tau \frac{dx}{x} f_\mathrm{a}\left( \frac{\tau}{x} \right)
f_\mathrm{b}(x)
\; \hat{\sigma}_\mathrm{ab\to BH}(\tau s) .
\end{equation}

\noindent
Since the trapped energy is a lower bound, the resulting cross section
is a lower bound.

\begin{figure}[tbp]
\begin{center}
\includegraphics[width=15cm]{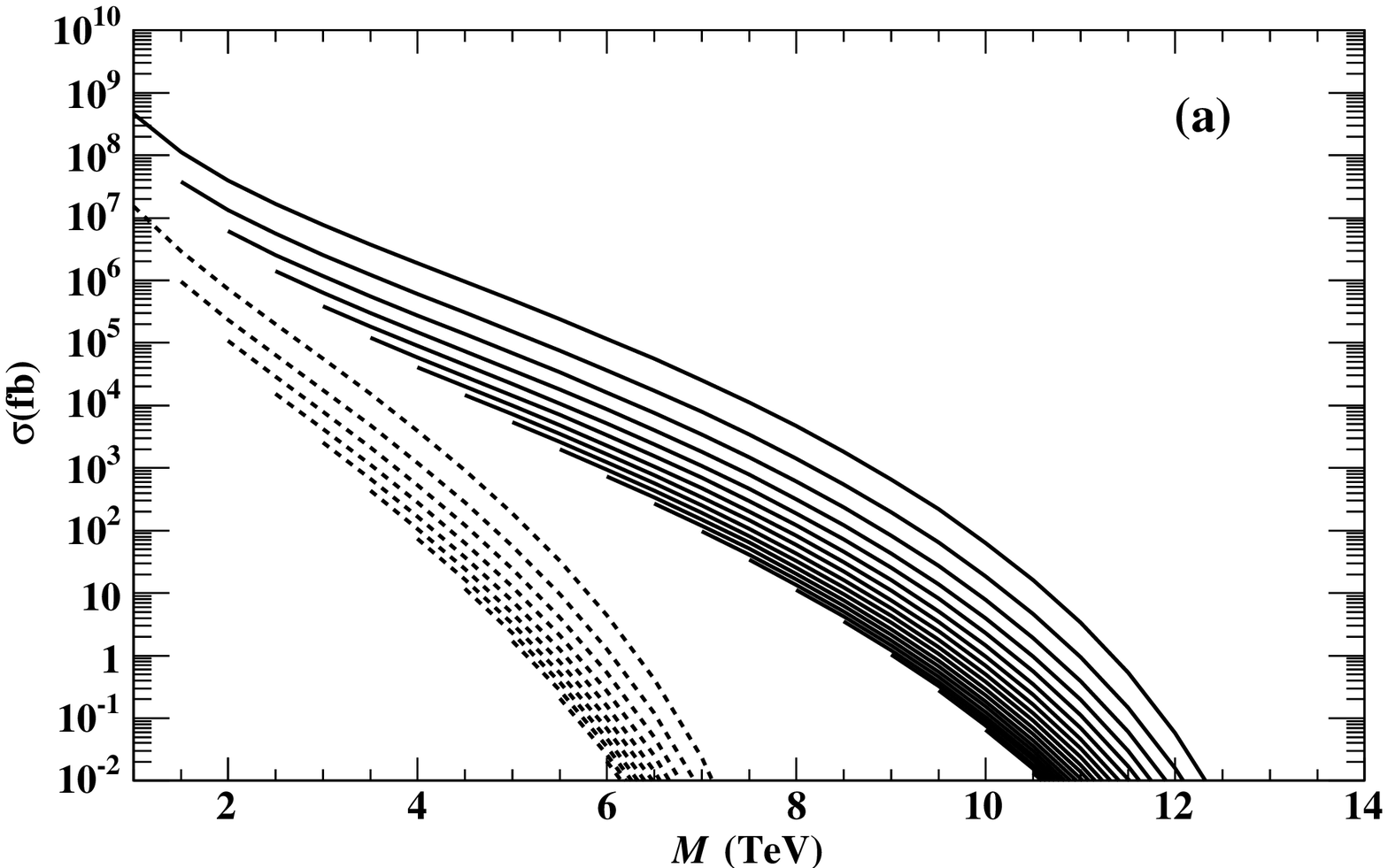}
\hfill
\includegraphics[width=15cm]{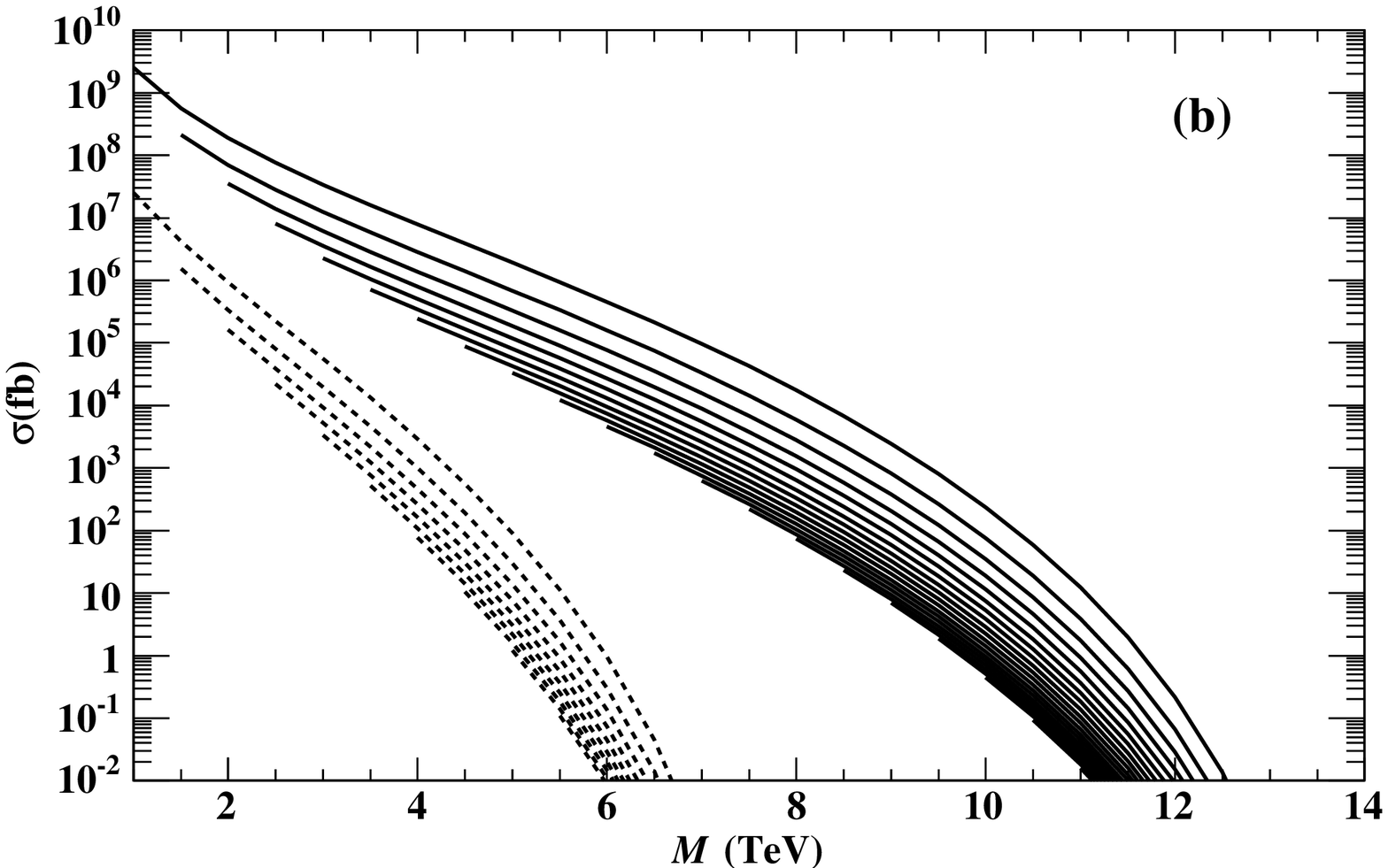}
\caption{\label{xsecpl} Cross section versus black hole mass. 
Solid curves semiclassical cross section and dashed curves trapped
surface cross section.   
Curves of same type for different Planck scales, 0.5~TeV top curves 
decreasing with increasing Planck scale in steps of 0.5~TeV. (a) $n=3$
and (b) $n=7$.}    
\end{center}
\end{figure}

\begin{figure}[tbp]
\begin{center}
\includegraphics[width=15cm]{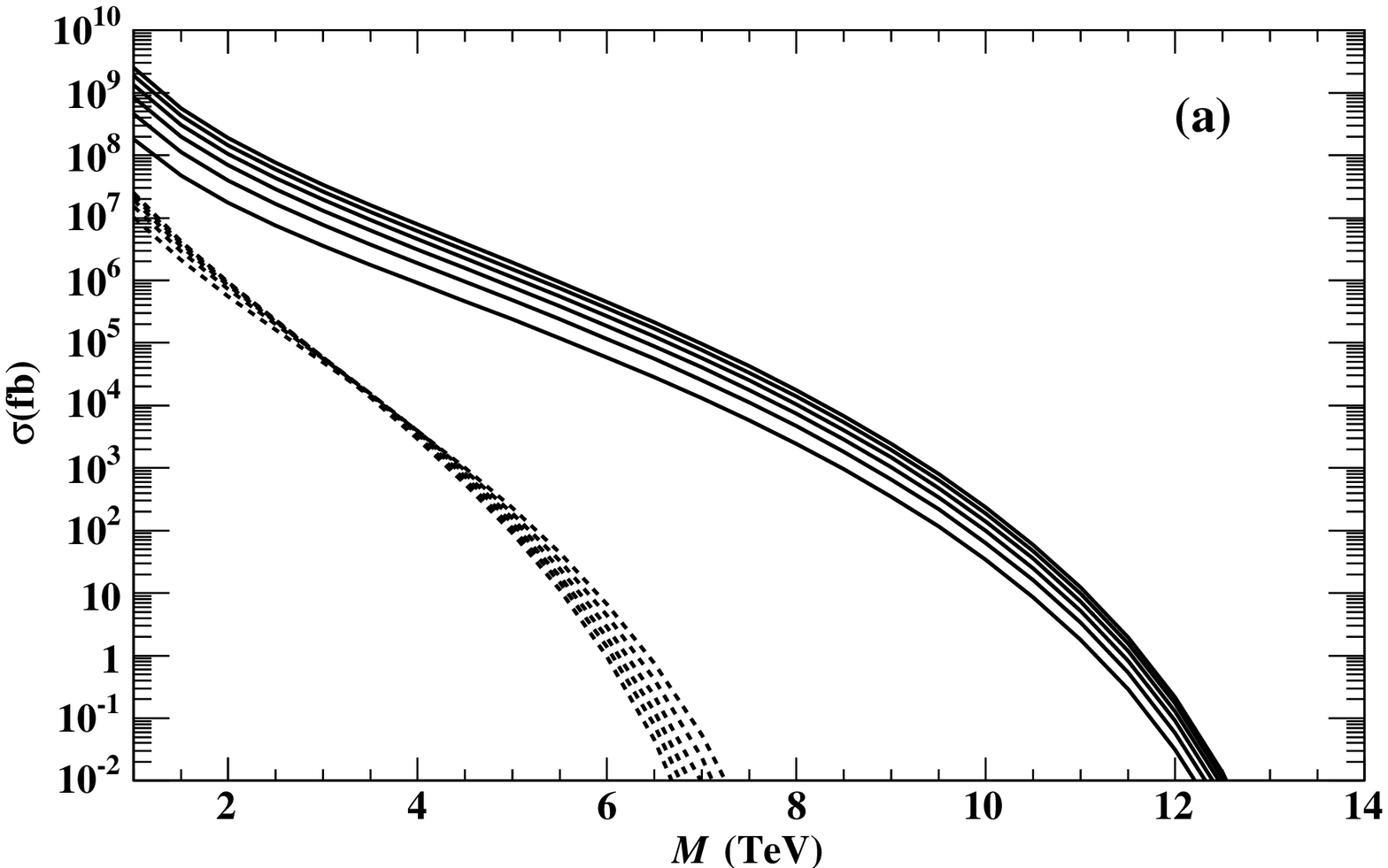}
\hfill
\includegraphics[width=15cm]{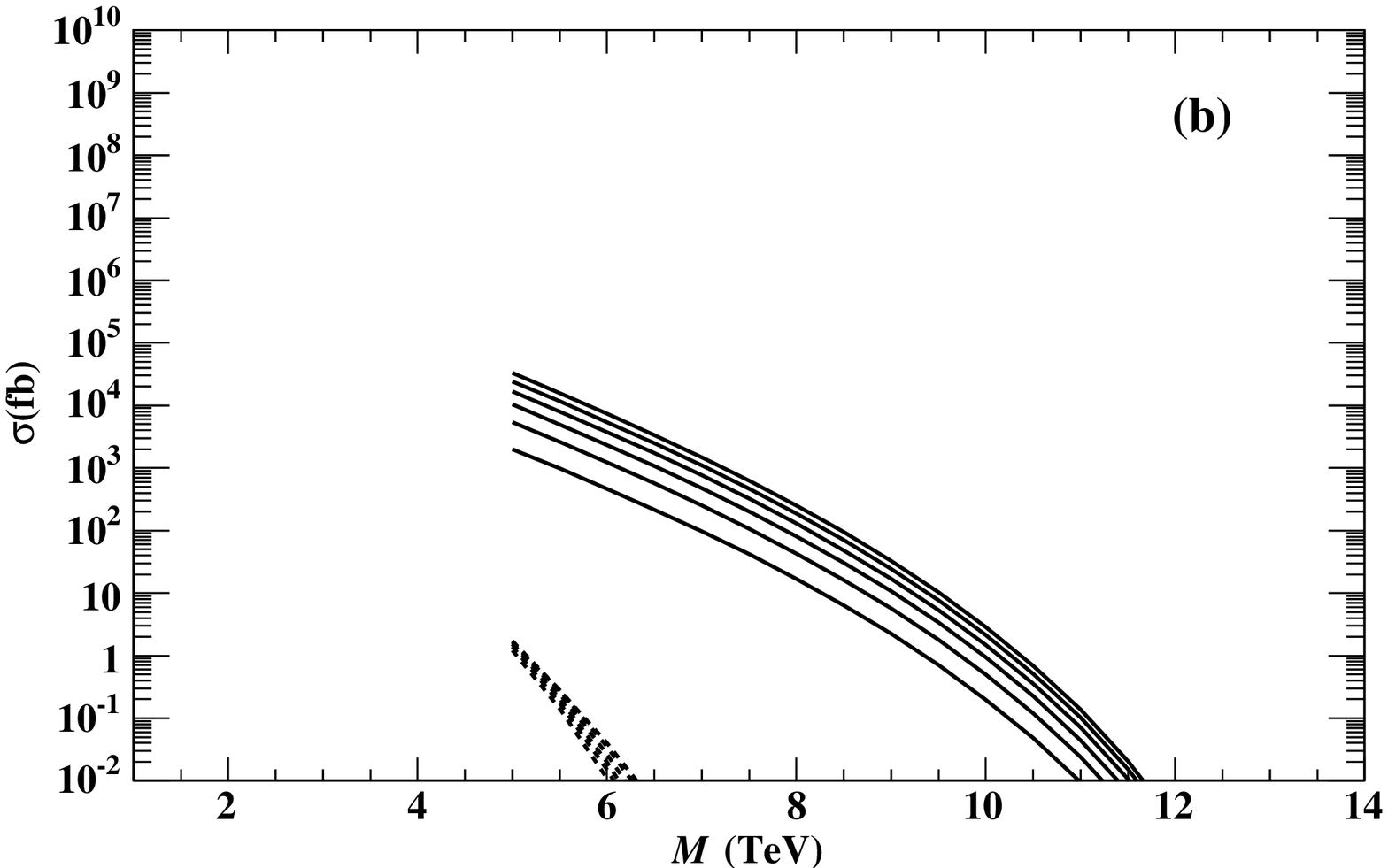}
\caption{\label{xsecn} Cross section versus black hole mass.
Solid curves semiclassical cross section and dashed curves trapped
surface cross section.   
Curves of same type for different number of extra dimensions, top curves
$n=2$ and bottom curves $n=7$. 
(a) $M_D=1$~TeV and (b) $M_D=5$~TeV.}
\end{center}
\end{figure}

Taking $x_\mathrm{min} = 1$, we obtain the families of cross section
curves shown in Figs.~\ref{xsecpl} and \ref{xsecn}.
The solid curves are for the semiclassical cross section calculated
using Eqs.~(\ref{eq7}), (\ref{eq8}), and (\ref{eq13}) with the form
factors $F_\mathrm{YR}$.  
We will henceforth refer to these curves as the semiclassical cross
section. 
The dashed lower curves are given by Eqs.~(\ref{eq13}) and (\ref{eq19})
with the form factors $F_\mathrm{YR}$. 
We will henceforth refer to these curves as the trapped surface (TS)
cross section.  
In Figs.~\ref{xsecpl}(a) and \ref{xsecpl}(b) the different curves of a given
type are for different Planck scales, starting from 0.5~TeV for the
top curve and decreasing with increasing Planck scale in steps of
0.5~TeV.
Figure~\ref{xsecpl}(a) is for $n=3$, while Fig.~\ref{xsecpl}(b) is for $n=7$. 
In Figs.~\ref{xsecn}(a) and \ref{xsecn}(b) the different curves of a given
type are for different numbers of extra dimensions, starting from
$n=2$ for the top curve and ending at $n=7$ for the bottom curve.
Figure~\ref{xsecn}(a) is for a Planck scale of 1~TeV, while
Fig.~\ref{xsecn}(b) is for a Planck scale of 5~TeV.  

The effect of non-trapped energy on the cross section is large because
the LHC energy is close to the threshold for black hole production and
lost energy limits the availability energy for the black hole.
The cross section curves show that there is less dependence on $n$ than
$M_D$.
This is because the $n$ dependence of the form factor tends to cancel
the $n$ dependence of the horizon radius.\cite{Webber}
It is reasonable to consider the semiclassical cross sections with form
factors greater than unity as loose upper bounds on the black hole cross
sections, which may increase by a factor of a few as the trapped-surface
cross sections increase.
We thus take the point of view that the black hole cross section lies
between the semiclassical and TS cross sections.
The difference can be several orders of magnitude.
The TS cross sections cut off at a mass above the trapped energy bounds
given by Fig.~\ref{trapped}.
Applying a cutoff $x_\mathrm{min} > 1$ will further restrict the range
of the TS cross sections, as well as the semiclassical cross sections.

\section{Lower Limits on the Planck Scale}

The cross sections in the previous section can be used to predict the
discovery limits for a given luminosity, and be used, in principle, to
extract the Planck scale and number of extra dimensions.
In the event of no detectable black hole signal, the cross sections call
also be used to set limits on the Planck scale and number of extra
dimensions.

\begin{figure}[tb]
\begin{center}
\includegraphics[width=15cm]{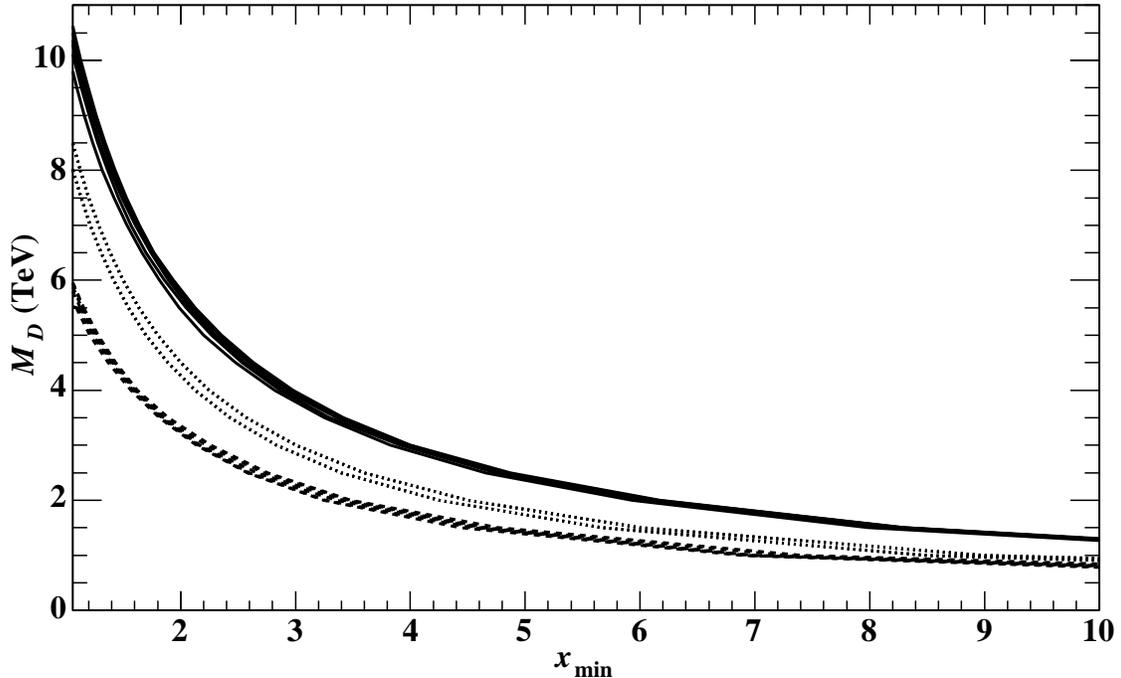}
\caption{\label{limit}Lower limits on Planck scale as function of cutoff
parameter.   
Solid curves semiclassical case.
Dashed curves trapped surface case.   
Dotted curves result of mass cutoff in trapped surface case.
Spread in curves of same type due to different $n$.} 
\end{center}
\end{figure}

We consider the scenario in which no black hole signal has been observed
after the accumulation of an integrated luminosity of 300~fb$^{-1}$ at
$\sqrt{s} = 14$~TeV.
Rather than study the different decay phases of the black hole and
estimate the detector's capabilities for measuring them, we assume a
perfect detector.
This will give the most optimistic limits possible.
Assuming a perfect experiment, the 95\% confidence-level upper limit on
the cross section is $10^{-2}$~fb.
Using this value of the cross section, we have extracted lower limits on
the Planck scale $M_D$ as a function of cutoff parameter
$x_\mathrm{min}$ for different values of the number of extra dimensions
$n$. 
The results are shown in Fig.~\ref{limit} for $n = 2$ to 7.
The solid curves were obtained from the semiclassical cross sections.  
The dashed curves were obtained from the trapped surface cross
section bounds. 
The dotted curves are a result of the mass cutoff in the trapped surface
cross sections.  
The dotted curves can be consider as the infinite luminosity case of the 
trapped surface predictions. 
The small spread in the different curves of a given type is due to
the different number of extra dimensions.

We can use Fig.~\ref{limit} to get a feel for how the different cross
section models affect the range of Planck-scale limits. 
For $x_\mathrm{min} = 5$, a lower limit of $M_D > 2.4$~TeV is obtained for
the semiclassical case and $M_D > 1.4$~TeV for the TS case.
The TS limit can be improved to $M_D > 1.7$~TeV with infinite luminosity.
Relaxing the cutoff criteria used to avoid quantum gravity effects to
$x_\mathrm{min} = 3$ gives a lower limit of $M_D > 3.8$~TeV for 
the semiclassical case and $M_D > 2.2$~TeV for the TS case.
The TS limit can be improved to $M_D > 2.8$~TeV with infinite luminosity.
There appears to be very little sensitivity to the limits on the Planck
scale due to the number of extra dimensions: less than a 3\% effect.

\section{Discussion}

The large difference in cross section between the models does not
translate into a large difference in the limits on $M_D$ because both
cross sections fall rapidly at low values of the cross section.
The limits on $M_D$ presented here are compatible with the discovery
limits that have been determined in previous
work.\cite{Giudice,Anchordoqui2,Anchordoqui3,Ahn2,Tanaka,Cambridge} 
Our limits might appear different due to the stringent requirements on
$x_\mathrm{min}$ and the different definition of $M_D$.
If one is willing to relax the requirement on $x_\mathrm{min}$ and risk
entering the quantum-gravity regime, than the differences between the two
models becomes significant.
This difference presumably still holds when the uncertainties in the
black hole decay and experimental effects are taken into account.

It is not possible to compare our results for the limits on $M_D$ with
existing experimental limits.
This would require estimating the uncertainties in the black hole decay
as well as including the experimental effects.
In any case, there are probably methods at the LHC that are more
sensitive to setting limits on $M_D$ than the direct search for black
holes.

After seven years, the validity of the naive geometric approximation for 
black hole cross section still stands. 
The best modifications we can make at this time is to include the form
factor in the last row of Table~\ref{t1}, which ranges from about 2 to 3
when the number of extra dimensions ranges from 2 to 7; not a very
significant modification when considering the other uncertainties.
However, classical general relativity has shown that not all the
available energy is trapped behind the horizon and lost energy should be 
accounted for.  
Unfortunately, the state of the calculations only allow us to
approximate the effects of non-trapped energy.

Ignoring the effects of non-trapped energy on the black hole cross
section could have a large affect on TeV-scale gravity studies.
This might modify the results of previous phenomenological studies,
which are sensitive to the black hole cross section relative to the QCD
cross section (see for example Ref.~\cite{Lonnblad1,Lonnblad2}).

The trapped-energy approach only gives a lower bound on the final mass 
of the black hole.  
In order to clarify the final mass, different methods such as the direct
study of gravitational wave emission are necessary.\cite{Cardoso1}  
The problem is extremely difficult because of the nonlinearity of
Einstein's equations, and because the high-energy collision of the two 
particles producing a black hole requires inclusion of nonlinear effects.
Cardoso \textit{et al}.\cite{Cardoso2,Cardoso3,Berti} have studied
gravitational radiation in linear perturbation theory of
higher-dimensional flat spacetime.    
The results are in agreement with the four dimensional estimate of
D'Eath and Payne.\cite{Eath1,Eath2,Eath3} 
However, the total energy decreases with the number of extra dimensions,
in disagreement with the estimate of Eardley and Giddings.\cite{Eardley}

Recently studies of gravitational radiation in the head-on collision of
two black holes in higher-dimensional spacetime have been made using a
close-slow approximation.\cite{Yoshino3,Yoshino4}  
This system can be regarded as a first simplification of particle
collisions. 
The results agree with the fully nonlinear analysis for four
dimensions. 

These perturbative approaches probably have an error of a factor of two
or more.
Since these studies are in the exploratory stage and have not yet been
performed as a function of impact parameter in higher dimensions, they
are currently of limited usefulness for making predictions at the LHC.

There are many uncertainties in our understanding of black hole
production in higher dimensional TeV-scale gravity. 
Reliable predictions of the cross section are not yet available.
We have examine the existing models to explore the different options for
filling in the gaps in our understanding.
In this way, we hope to be better prepared to confront the possibility
of black hole production at the LHC.

\textit{Note added.} While this work was being finished, we learned that
some aspects of the topics addressed in this paper were also under 
consideration by another group; that work has recently appearance
in Ref.~\cite{Cavaglia,Ahn3}. 

\section*{Acknowledgments}

This work was supported in part by the Natural Sciences and Engineering
Research Council of Canada.

\appendix
\section{\boldmath{$D$}-Dimensional Black Hole Horizon Radius}

Black hole Kerr solutions to Einstein's equations have been obtained in
asymptotically flat higher dimensional spacetime.\cite{Myers}
For distances much smaller than size of the extra dimensions, the event
horizon radius for a $D$-dimensional spinning black hole is given by  

\begin{equation}
r_h^{D-5} \left[ r_h^2 + \frac{(D-2)^2J^2}{4M^2} \right] = \frac{16\pi
G_D M}{(D-2)\Omega_{D-2}} ,
\end{equation}

\noindent
where $J$ is the four-dimensional angular momentum, $M$ is the mass of
the black hole, and $\Omega_{D-2}$ is the area of a unit ($D-2$)-sphere, 
given by  

\begin{equation}
\Omega_{D-2} = \frac{2\pi^{(D-1)/2}}{\Gamma\left( \frac{D-1}{2} \right)} .
\end{equation}

\noindent
A common way of writing the horizon radius is

\begin{equation}
r_\mathrm{h} = \left[ \frac{1}{1 + a^2} \right]^\frac{1}{D-3} R_\mathrm{S} ,
\end{equation}

\noindent
where 

\begin{equation}
a = \frac{(D-2)J}{2Mr_\mathrm{h}} 
\end{equation}

\noindent
is a rotation parameter depending on the horizon radius, and 

\begin{equation}
R_\mathrm{S} = \left[ G_D M \left( 2^3 \pi^{(3-D)/2} \frac{\Gamma
\left( \frac{D-1}{2} \right) }{(D-2)} \right) \right]^\frac{1}{D-3}
\end{equation}

\noindent
is the spinless (Schwarzschild) horizon radius.  
When $J\to 0$, $r_\mathrm{h} \to R_\mathrm{S}$.

To express the $D$-dimensional Newton constant in terms of the Planck
scale, we must choose a convention. 
Using the PDG convention (our convention)

\begin{equation} \label{eq26}
M_D^{D-2} = \frac{(2\pi)^{D-4}}{8\pi G_D} = \frac{(2\pi)^n}{8\pi G_D} ,
\end{equation}

\noindent
the horizon radius is

\begin{eqnarray} 
R_\mathrm{S} & = & \frac{1}{M_D} \left[
\frac{M}{M_D} \left( 2^{D-4} \pi^{(D-7)/2}\frac{\Gamma\left(
\frac{D-1}{2} \right)}{D-2} \right) \right]^{\frac{1}{D-3}} \label{eq27s}\\
&= & \frac{1}{M_D} \left[ \frac{M}{M_D} \left( 2^n
  \sqrt{\pi}^{(n-3)}\frac{\Gamma\left( \frac{n+3}{2} \right)}{n+2} \right)
  \right]^{\frac{1}{n+1}} . \label{eq27b} \nonumber \\
\end{eqnarray}

\noindent
Using the convention of Dimopoulos and Landsberg\cite{Dimopoulos1} 

\begin{equation} \label{eq28}
M_\mathrm{DL}^{D-2} = \frac{1}{G_D} ,
\end{equation}

\noindent
the horizon radius is

\begin{eqnarray} 
R_\mathrm{S} & = & \frac{1}{\sqrt{\pi}M_\mathrm{DL}} \left[
\frac{M}{M_\mathrm{DL}} \left( \frac{8 \Gamma\left( \frac{D-1}{2}
\right)}{D-2} \right) \right]^{\frac{1}{D-3}} \label{eq29a} \\
& = & \frac{1}{M_\mathrm{DL}} \left[
\frac{M}{M_\mathrm{DL}} \left( 2^3 \sqrt{\pi}^{(-n-1)} \frac{
\Gamma\left( \frac{n+3}{2} \right)}{n+2} \right) \right]^{\frac{1}{n+1}}
. \label{eq29b} \nonumber \\
\end{eqnarray}

Although the Dimopoulos and Landsberg convention results in a simpler
expression for the Planck scale Eq.~(\ref{eq28}) and the cross section
Eq.~(\ref{eq29a}), experimental limits are most often set on the Planck
scale in the PDG convention Eq.~(\ref{eq26}).

Setting the black hole mass equal in the two conventions for the horizon
radii Eq.~(\ref{eq27b}) and Eq.~(\ref{eq29b}), the ratio of the cross
sections can be written as\cite{Rizzo1} 

\begin{equation}
\frac{\hat{\sigma}}{\hat{\sigma}_\mathrm{DL}} =
\left[ \frac{(2\pi)^n}{8\pi} \right]^\frac{2}{n+1} \left[
\frac{M_\mathrm{DL}^2}{M_D^2} \right]^\frac{n+2}{n+1} .
\end{equation}

\noindent
For $M_\mathrm{DL} = M_D$, the ratio of the cross sections is 1.4 for
$n=2$ and 11.1 for $n=7$. 


\begin{thebibliography}{00}

\bibitem{ADD1}
N. Arkani-Hamed, S. Dimopoulos and G. Dvali,
``The hierarchy problem and new dimensions at a millimeter,''
\textit{Phys. Lett.} \textbf{B429}, 263 (1998); arXiv:hep-ph/9803315.

\bibitem{ADD2}
I. Antoniadis, N. Arkani-Hamed, S. Dimopoulos and G. Dvali,
``New dimensions at a millimeter to a fermi and superstings at a TeV,'' 
\textit{Phys. Lett.} \textbf{B436}, 257 (1998); arXiv:hep-ph/9804398.

\bibitem{ADD3}
N. Arkani-Hamed, S. Dimopoulos and G. Dvali,
``Phenomenology, astrophysics, and cosmology of theories with
submillimeter dimensions and TeV scale quantum gravity,'' 
\textit{Phys. Rev.} \textbf{D59}, 086004 (1999); arXiv:hep-ph/9807344.

\bibitem{RS1}
L. Randall and R. Sundrum,
``Large Mass Hierarchy from a Small Extra Dimension,''
\textit{Phys. Rev. Lett.} \textbf{83}, 3370 (1999); arXiv:hep-ph/9905221.

\bibitem{RS2}
L. Randall and R. Sundrum,
``An Alternative to Compactification,''
\textit{Phys. Rev. Lett.} \textbf{83}, 4690 (1999); arXiv:hep-th/9906064.

\bibitem{Banks}
T. Banks and W. Fischler,
A Model for High Energy Scattering in Quantum Gravity,
arXiv:hep-th/9906038.

\bibitem{Giddings1}
S.~B. Giddings and S. Thomas, 
``High energy colliders as black hole factories: The end of short
distance physics,'' 
\textit{Phys. Rev.} \textbf{D65}, 056010 (2002); arXiv:hep-ph/0106219. 

\bibitem{Dimopoulos1}
S. Dimopoulos and G. Landsberg, 
``Black Holes at the Large Hadron Collider,'' 
\textit{Phys. Rev. Lett.} \textbf{87}, 161602 (2001); arXiv:hep-ph/0106295.

\bibitem{Voloshin1}
M.~B. Voloshin, 
``Semiclassical suppression of black hole production in particle
collisions,'' 
\textit{Phys. Lett.} \textbf{B518}, 137 (2001); arXiv:hep-ph/0107119.

\bibitem{Voloshin2}
M.~B. Voloshin, 
``More remarks on suppression of large black hole production in particle
collisions,''
\textit{Phys. Lett.} \textbf{B524}, 376 (2001); arXiv:hep-ph/0111099.

\bibitem{Solodukhin}
S.~N. Solodukhin,
``Classical and quantum cross-section for black hole production in
particle collisions", 
\textit{Phys. Lett.} \textbf{B533}, 153 (2002); arXiv:hep-ph/0201248.

\bibitem{Jevicki}
A. Jevicki and J. Thaler,
``Dynamics of black hole formation in an exactly solvable model,''
\textit{Phys. Rev.} \textbf{D66}, 024041 (2002); arXiv:hep-th/0203172.

\bibitem{Rychkov04}
V.~S. Rychkov,
``Topics in Black Hole Production,''
Carg{\`e}se Summer School, June 7-19, 2004, 363-369; arXiv:th/0410295. 

\bibitem{Park}
S.~C. Park and H.~S. Song,
``Production of Spinning Black Holes at colliders,''
\textit{J. Korean Phys. Soc.} \textbf{43}, 33 (2003); arXiv:hep-ph/0111069.

\bibitem{Kotwal}
A.~V. Kotwal and C. Hays,
``Production and decay of spinning black holes at colliders and tests of
black hole dynamics,''
\textit{Phys. Rev.} \textbf{D66}, 116005 (2002); arXiv:hep-ph/0206055.

\bibitem{Anchordoqui1}
L. Anchordoqui, J.~L. Feng, H. Goldberg and A.~D. Shapere,
``Black holes from cosmic rays: Probes of extra dimensions and new
limits on TeV-scale gravity,''
\textit{Phys. Rev.} \textbf{D65}, 124027 (2002); arXiv:hep-ph/0112247.

\bibitem{Ida}
D. Ida, K. ya Oda and S.~C. Park, 
``Rotating black holes at future colliders: Greybody factors for brane
fields,'' 
\textit{Phys. Rev.} \textbf{D67}, 064025 (2003);
Erratum-ibd. \textbf{D69}, 049901 (2004); arXiv:hep-th/0212108. 

\bibitem{Eardley}
D.~M. Eardley and S.~B. Giddings, 
``Classical black hole production in high-energy collisions,'' 
\textit{Phys. Rev.} \textbf{D66}, 044011 (2002); arXiv:gr-qc/0201034.

\bibitem{Yoshino1}
H. Yoshino and Y. Nambu, 
``Black hole formation in the grazing collision of high-energy
particles,'' 
\textit{Phys. Rev.} \textbf{D67}, 024009 (2003); arXiv:gr-qc/0209003.

\bibitem{Kohlprath}
E. Kohlprath and G. Veneziano,
``Black holes from high-energy beam-beam collisions,''
\textit{J. High Energy Phys.} \textbf{0206}, 057 (2002); arXiv:gr-qc/0203093.

\bibitem{Yoshino2}
H. Yoshino and V.~S. Rychkov, 
``Improved analysis of black hole formation in high-energy particle
collisions,'' 
\textit{Phys. Rev.} \textbf{D71}, 104028 (2005); arXiv:hep-th/0503171.

\bibitem{Mann}
H. Yoshino and R.~B. Mann,
``Black hole formation in the head-on collision of ultarelativistic
charges,''
\textit{Phys. Rev.} \textbf{D74}, 044003 (2006); arXiv:gr-qc/0605131.

\bibitem{Rychkov}
V.~S. Rychkov,
``Black hole production in particle collisions and higher curvature
gravity,''
\textit{Phys. Rev.} \textbf{D70}, 044003 (2004); arXiv:hep-ph/041116.

\bibitem{Giddings2}
S.~B. Giddings and V.~S. Rychkov,
``Black holes from colliding wavepackets,''
\textit{Phys. Rev.} \textbf{D70}, 104026 (2004); arXiv:hep-th/0409131.

\bibitem{Aliev}
A.~N. Aliev and A.~E. G\"{u}mr\"{u}k\c{c}\"{u}o\u{g}lu,
``Charged rotating black holes on a 3-brane",
\textit{Phys. Rev.} \textbf{D71}, 104027 (2005); arXiv:hep-ph/0607027.

\bibitem{Rocha}
R. Rocha and C.~H. Coimbra-Ara{\'u}jo,
``Extra dimensions in LHC via mini-black holes: effective Kerr-Newman
brane-world effects'',
\textit{Phys. Rev.} \textbf{D74}, 055006 (2006); arXiv:hep-ph/0607027.

\bibitem{PDG}
Particle Data Group (W.-M. Yoa \textit{et al}.),
\textit{J. Phys. G} \textbf{33}, 1165 (2006).

\bibitem{Giudice}
G.~F. Giudice, R. Rattazzi and J.~D. Wells,
``Transplanckian Collisions at the LHC and Beyond,''
\textit{Nucl. Phys.} \textbf{B630}, 293 (2002); arXiv:hep-ph/0112161.

\bibitem{Voloshin3}
M.~B. Voloshin,
Erratum to: ``More remarks on suppression of large black hole production
in particle collisions'' [Phys. Lett. B 524 (2002) 376],
\textit{Phys. Lett.} \textbf{B605}, 426 (2005).

\bibitem{Rizzo1}
T.~G. Rizzo, 
``Black Hole Production Rates at the LHC: Still Large,''
eConf C010630, P339 (2001); arXiv:hep-ph/0111230. 

\bibitem{Emparan}
R. Emparan, M. Masip and R. Rattazzi,
``Cosmic rays as probes of large extra dimensions and TeV gravity,''
\textit{Phys. Rev.} \textbf{D65}, 064023 (2002); arXiv:hep-ph/0109287.

\bibitem{Pumplin}
J. Pumplin, D.~R. Stump, J. Huston, H.~L. Lai, P. Nadolsky and W.~K. Tung,
``New Generation of Parton Distributions with Uncertainties from Global
QCD Analysis,''
\textit{J. High Energy Phys.} \textbf{0407}, 012 (2002); arXiv:hep-ph/0201195.

\bibitem{LHAPDF}
LHAPDF the Les Houches Accord PDF Interface, Version 5.2.2, maintained
by M. Whalley; http://hepforge.cedar.ac.uk/lhapdf/.

\bibitem{Myers}
R.~C. Myers and M.~J. Perry, 
``Black Holes in Higher Dimensional Space-Times,'' 
\textit{Ann. Phys.} (N.Y.) \textbf{172}, 304 (1986).

\bibitem{Koch} 
B. Koch, M. Bleicher and S. Hossenfelder, 
``Black hole remnants at the LHC,''
\textit{J. High Energy Phys.} \textbf{0510}, 053 (2005); arXiv:hep-ph/0507138. 

\bibitem{Rizzo2} 
T.~G. Rizzo,
TeV-Scale Black Holes in Warped Higher-Curvature Gravity,
arXiv:hep-ph/0510420.

\bibitem{Rizzo3} 
T.~G. Rizzo,
Noncommutative Inspired Black Holes in Extra Dimensions,
\textit{J. High Energy Phys.} \textbf{0609}, 021 (2006); 
arXiv:hep-ph/0606051.

\bibitem{Dimopoulos2} 
S. Dimopoulos and R. Emparan,
``String balls at the LHC and beyond,''
\textit{Phys. Lett.} \textbf{B526}, 393 (2002); arXiv:hep-ph/0108060.

\bibitem{Casadio} 
R. Casadio and B. Harms,
``Can black holes and naked singularities be detected in accelerators?,''
\textit{Int. J. Mod. Phys.} \textbf{A17}, 4635 (2002); arXiv:hep-th/0110255.

\bibitem{Ahn1}
E.-J. Ahn,  M. Cavagli\`{a} and A.~V. Olinto,
``Brane factories", 
\textit{Phys. Lett.} \textbf{B551}, 1 (2003); arXiv:hep-th/0201042.

\bibitem{Cheung1} 
K. Cheung and C.-H. Chou, 
``$p$-brane production in Fat brane or Universal extra dimensions
scenario,'' 
\textit{Phys. Rev.} \textbf{D66}, 036008 (2002); arXiv:hep-ph/0205284.   

\bibitem{Cheung2} 
K. Cheung, 
``Black hole, string ball, and $p$-brane production at hadronic
supercolliders,'' 
\textit{Phys. Rev.} \textbf{D66}, 036007 (2002); arXiv:hep-ph/0205033.

\bibitem{Arkani1}
N. Arkani-Hamed and M. Schmaltz,
``Hierarchies without Symmetries from Extra Dimensions,''
\textit{Phys. Rev.} \textbf{D61}, 033005 (2000); arXiv:hep-ph/9903417.

\bibitem{Arkani2}
N. Arkani-Hamed, Y. Grossman and M Schmaltz,
``Split Fermions in Extra Dimensions and Exponentially Small
Cross-Sections at Future Colliders,''
\textit{Phys. Rev.} \textbf{D61}, 115004 (2000); arXiv:hep-ph/9909411.

\bibitem{Dai}
D.-C. Dai, G.~D. Starkman and D. Stojkovic,
``Production of black holes and their angular momentum distribution in
models with split fermions,'' 
\textit{Phys. Rev.} \textbf{D73}, 104037 (2006); arXiv:hep-ph/0605085.

\bibitem{Hossenfelder}
S. Hossenfelder, 
``Suppressed black hole production from minimal length,'' 
\textit{Phys. Lett.} \textbf{B598}, 92 (2004); arXiv:hep-ph/0404232.

\bibitem{Anchordoqui2}
L. Anchordoqui, J.~L. Feng, H. Goldberg and A.~D. Shapere,
``Inelastic black hole production and large extra dimensions,''
\textit{Phys. Lett.} \textbf{B594}, 363 (2004); arXiv:hep-ph/0311365.

\bibitem{Gibbons}
G.~W. Gibbons and S.~W. Hawking,
``Action integrals and partition functions in quantum gravity,''
\textit{Phys. Rev.} \textbf{D15}, 2752 (1977).

\bibitem{Aichelburg}
P.~C. Aichelburg and R.~U. Sexl,
``On the Gravitational Field of a Massless Particle,''
\textit{Gen. Rel. Grav.} \textbf{2}, 303 (1971).

\bibitem{Penrose}
R. Penrose (unpublished, reported in Ref.~\cite{Eath3})

\bibitem{Eath1}
P.~D. D'Eath and P.~N. Payne,
``Gravitational radiation in black-hole collisions at the speed of
light.
I. Perturbation treatment of the axisymmetric collision,''
\textit{Phys. Rev.} \textbf{D46}, 658 (1992). 

\bibitem{Eath2}
P.~D. D'Eath and P.~N. Payne,
``Gravitational radiation in black-hole collisions at the speed of
light.
II. Reduction to two independent variables and calculation of the
second-order news function,''
\textit{Phys. Rev.} \textbf{D46}, 675 (1992).

\bibitem{Eath3}
P.~D. D'Eath and P.~N. Payne,
``Gravitational radiation in black-hole collisions at the speed of
light.
III. Results and conclusions,'' 
\textit{Phys. Rev.} \textbf{D46}, 694 (1992).

\bibitem{Cardoso1}
V. Cardoso, E. Berti and M. Cavagli{\`a},
``What we (don't) know about black hole formation in high-energy
collisions,'' 
\textit{Class. Quant. Grav.} \textbf{22}, L61 (2005); arXiv:hep-ph/0505125.

\bibitem{Anchordoqui3}
L. Anchordoqui, J.~L. Feng, H. Goldberg and A.~D. Shapere,
``Updated limits on TeV-scale gravity from the absence of neutrino
cosmic ray showers mediated by black holes,''
\textit{Phys. Rev.} \textbf{D68}, 104025 (2003); arXiv:hep-ph/0307228.

\bibitem{Webber}
B. Webber,
``Black Holes at Accelerators,''
eConf C0507252, T030 (2005); arXiv:hep-ph/0511128.

\bibitem{Ahn2}
E.-J. Ahn M. Cavagli\`{a} and A.~V. Olinto,
``Uncertainties in limits on TeV-gravity from neutrino-induced showers," 
\textit{Astropart. Phys.} \textbf{22}, 377 (2005); arXiv:hep-ph/0312249.

\bibitem{Tanaka}
J. Tanaka, T. Yamamura, S. Asai and J. Kanzaki,
``Study of black holes with the ATLAS detector at the LHC,''
\textit{Eur. Phys. J.} \textbf{C41}, s02 19 (2005); arXiv:hep-ph/0411095.

\bibitem{Cambridge}
C.~M. Harris, M.~J. Palmer, M.~A. Parker, P. Richardson, A. Sabetfakhri and
B.~R. Webber,
``Exploring higher dimensional black holes at the Large Hadron
Collider,''
\textit{J. High Energy Phys.} \textbf{0505}, 053 (2005); arXiv:hep-ph/0411022.

\bibitem{Lonnblad1}
L. L{\"o}nnblad, M. Sj{\"o}dahl and T. \AA esson,
``QCD-suppression by Black Hole Production at the LHC,''
\textit{J. High Energy Phys.} \textbf{0509}, 019 (2005); arXiv:hep-ph/0505181.

\bibitem{Lonnblad2}
L. L{\"o}nnblad and M. Sj{\"o}dahl,
``Classical and Non-Classical ADD-phenomenology with high-$E_\perp$ jet
observables at collider experiments,''
\textit{J. High Energy Phys.} \textbf{0610}, 088 (2006); arXiv:hep-ph/0608210.

\bibitem{Cardoso2}
V. Cardoso, O. J. C. Dias and P. S. Lemos,
``Gravitational radiation in $D$-dimensional spacetimes,"
\textit{Phys. Rev.} \textbf{D67}, 064016 (2003); arXiv:hep-th/0212168.

\bibitem{Cardoso3}
V. Cardoso, P. S. Lemos and S. Yoshida,
``Electromagnetic radiation from collisions at almost the speed of light:
An extremely relativistic charged particle falling into a Schwarzschild
black hole,"
\textit{Phys. Rev.} \textbf{D68}, 084011 (2003); arXiv:gr-qc/0307104.

\bibitem{Berti}
E. Berti, M. Cavagli{\`a} and L. Gualtieri,
``Gravitational energy loss in high energy particle collisions:
Ultrarelativistic plunge into a multidimensional black hole,"
\textit{Phys. Rev.} \textbf{D69}, 124011 (2004); arXiv:hep-th/0309203.

\bibitem{Yoshino3}
H. Yoshino, T. Shiromizu and M. Shibata,
``Close-limit analysis for head-on collision of two black holes in higher
dimensions: Brill-Lindquist initial data,"
\textit{Phys. Rev.} \textbf{D72}, 084010 (2005); arXiv:gr-qc/0508063.

\bibitem{Yoshino4}
H. Yoshino, T. Shiromizu and M. Shibata,
``Close-slow analysis for head-on collision of two black holes in higher
dimensions: Bowen-York initial data,'' 
\textit{Phys. Rev.} \textbf{D74}, 124022 (2006); arXiv:gr-qc/0610110.

\bibitem{Cavaglia}
M. Cavagli{\`a}, R. Godang, L. Cremaldi and D. Summers,
Catfish: A Monte Carlo simulator for black holes at the LHC,
arXiv:hep-ph/0609001.

\bibitem{Ahn3}
E.-J. Ahn and M. Cavagli\`{a},
``Simulations of black hole air showers in cosmic ray detectors",
\textit{Phys. Rev.} \textbf{D73}, 042002 (2006); arXiv:hep-ph/0511159.

\end{thebibliography}

\end{document}